\documentclass[12pt,draftcls, onecolumn]{IEEEtran}

\usepackage{times}
\usepackage{amsmath,dsfont}
\usepackage{amssymb,amsthm}
\usepackage{epsfig,verbatim}
\usepackage{subfigure}
\usepackage{setspace}
\usepackage{color}
\usepackage{cite}
\usepackage{epstopdf}
\usepackage{graphics}
\usepackage{accents}
\usepackage{acronym}
\usepackage[bookmarks,colorlinks]{hyperref}

\newtheorem{definition}{Definition}
\newtheorem{theorem}{Theorem}
\newtheorem*{theorem*}{Theorem}
\newtheorem{corollary}{Corollary}
\newtheorem{proposition}{Proposition}
\newtheorem{lemma}{Lemma}
\newtheorem{remark}{Comment}

\newcommand{\E}{\mathds{E}}

\newcommand{\myMat}[1]{\mathsf{#1}}

\newcommand{\Rs}{R_{\rm{c}}}
\newcommand{\Xin}{{\bf{X}}}
\newcommand{\xin}{{\bf{x}}}
\newcommand{\Yi}{{\bf{Y}}}
\newcommand{\yi}{{\bf{y}}}
\newcommand{\Wi}{{\bf{W}}}
\newcommand{\wi}{{\bf{w}}}
\newcommand{\Ui}{{\bf{U}}}
\newcommand{\ui}{{\bf{u}}}
\newcommand{\Mem}{m}
\newcommand{\Pdf}[1]{p\left( #1\right) }
\newcommand{\Hi}{\myMat{G}}
\newcommand{\Fi}{\myMat{F}}
\newcommand{\PCst}{P}
\newcommand{\CWi}{{\myMat{C}_{\Wi}}}

\newcommand{\HWi}{\bar{H}_{\Wi}}
\newcommand{\HUi}{\bar{H}_{\Ui}}
\newcommand{\Hfd}{\myMat{G}'}
\newcommand{\Ffd}{\myMat{F}'}
\newcommand{\CWfd}{\myMat{C}_{\Wi}'}
\newcommand{\HWiG}{\bar{H}_{G,\Wi}}

\newcommand{\Capacity}{C_{\rm L}}
\newcommand{\TCapacity}{C_{\rm DCD}}
\newcommand{\CapGauss}{C_{\rm G}}

\newcommand{\Xeq}{\tilde{\bf X}}

\newcommand{\Yeq}{\tilde{\bf Y}}

\newcommand{\Weq}{\tilde{\bf W}}
\newcommand{\Weqs}{\tilde{W}}
\newcommand{\Ueq}{\tilde{\bf U}}

\newcommand{\Heq}{\tilde{\myMat{G}}}
\newcommand{\PCsteq}{\tilde{P}}
\newcommand{\Memeq}{\tilde{m}}
\newcommand{\Capeq}{C_{\rm PLC}}
\newcommand{\Cov}{{\rm Cov}}
\newcommand{\BB}{_{\rm DCD}}

\newcommand{\nt}{n_{\rm t}}
\newcommand{\nr}{n_{\rm r}}
\newcommand{\tnt}{\tilde{n}_{\rm t}}
\newcommand{\tnr}{\tilde{n}_{\rm r}}

\newcommand{\CWiBB}{{\myMat{C}_{\Wi\BB}}}
\newcommand{\HWiBB}{\bar{H}_{\Wi\BB}}
\newcommand{\CWfdBB}{\myMat{C}_{\Wi\BB}'}
\newcommand{\HWiGBB}{\bar{H}_{G,\Wi\BB}}
\newcommand{\CapGaussBB}{C_{{\rm DCD},G}}

\newcommand{\ieq}{{\tilde i}}
\newcommand{\teq}{{\tilde \tau}}
\newcommand{\lequ}{{\tilde l}}
\newcommand{\PerCh}{\tilde{p}_{\mathsf{G}}}
\newcommand{\PerN}{\tilde{p}_{\bf W}}
\newcommand{\Per}{\tilde{p}}
\newcommand{\PerSet}{\mathcal{P}}

\newcommand{\mvn}[4]{f_{G_{#4}}\big(#1;#2,#3\big)}

\newcommand{\TDelta}{{{\Delta'}}}
\newcommand{\Tlambda}{\lambda'}
\newcommand{\TlambdaN}{\alpha'}

\newcommand{\Nakagami}[2]{{\mathcal{KG}}\left(#1,#2 \right) }
\newcommand{\CNakagami}[2]{{\mathcal{CKG}}\left(#1,#2 \right) }

\newcommand{\pdf}[2]{f_{#1}\left(#2 \right)}

\newcommand{\prior}[1]{{\gamma_{#1}}}
\newcommand{\mean}[1]{{{\bf m}_{#1}}}
\newcommand{\var}[1]{{\myMat{C}_{#1}}}
\newcommand{\nG}{n_{\rm G}}
\newcommand{\mN}{\mathcal{N}}

\newcommand{\WW}{W}
\newcommand{\ww}{w}
\newcommand{\WWr}{W_{\rm R}}
\newcommand{\wwr}{w_{\rm R}}
\newcommand{\WWi}{W_{\rm I}}
\newcommand{\wwi}{w_{\rm I}}

\newcommand{\Complex}{^{\rm C}}

\newif\ifcomments
\definecolor{CmtColor}{rgb}{0,0.6,1}

\long\def\symbolfootnote[#1]#2{\begingroup\def\thefootnote{\fnsymbol{footnote}}\footnote[#1]{#2}\endgroup}

\acrodef{bb}[BB]{broadband}
\acrodef{dt}[DT]{discrete-time}
\acrodef{ct}[CT]{continuous-time}
\acrodef{hsr}[HSR]{harmonic series representation}
\acrodef{dcd}[DCD]{decimated components decomposition}
\acrodef{ctf}[CTF]{channel transfer function}
\acrodef{cir}[CIR]{channel impulse response}
\acrodef{awgn}[AWGN]{additive white Gaussian noise}
\acrodef{acgn}[ACGN]{additive cyclostationary Gaussian noise}
\acrodef{wss}[WSS]{wide-sense stationary}
\acrodef{pdf}[PDF]{probability density function}
\acrodef{ofdm}[OFDM]{orthogonal frequency division multiplexing}
\acrodef{cp}[CP]{cyclic prefix}
\acrodef{mimo}[MIMO]{multiple input-multiple output}
\acrodef{lgmc}[LNGMC]{linear non-Gaussian MIMO channel}
\acrodef{nlgmc}[$n$-MNGMC]{$n$-block memoryless non-Gaussian MIMO channel}
\acrodef{lti}[LTI]{linear time-invariant}
\acrodef{lptv}[LPTV]{linear periodically time-varying}
\acrodef{isi}[ISI]{intersymbol interference}
\acrodef{snr}[SNR]{signal-to-noise ratio}
\acrodef{gm}[GM]{Gaussian mixture}
\acrodef{rv}[RV]{random variable}
\acrodef{rms}[RMS]{root mean-square}
\acrodef{cn}[CN]{complex normal}
\acrodef{dft}[DFT]{discrete Fourier transform}
\acrodef{dtft}[DTFT]{discrete time Fourier transform}
\acrodef{psd}[PSD]{power spectral density}
\acrodef{wscs}[WSCS]{wide-sense cyclostationary}
\acrodef{kld}[KLD]{Kullback-Leibler divergence}
\acrodef{plc}[PLC]{Power line communications}

 \IEEEoverridecommandlockouts
\title{On the Capacity of MIMO Broadband Power Line Communications Channels
}

\vspace{-0.5cm}

\author{
\IEEEauthorblockN{\vspace{-0.2cm} Nir Shlezinger, Roee Shaked, and Ron Dabora\\
}

\thanks{Nir Shlezinger is with the Faculty of Electrical Engineering, Technion, Haifa, Israel (e-mail:	nirshlezinge@technion.ac.il). Roee Shaked and Ron Dabora are with the department of ECE, Ben-Gurion University, Be'er-Sheva, Israel (e-mail:shroee@post.bgu.ac.il; ron@ee.bgu.ac.il). This work was supported by the Israel Science Foundation under Grant 1685/16.}

\vspace{-1.0cm}

}

\vspace{-0.95cm}

\begin{document}

\maketitle
\pagestyle{plain}
\thispagestyle{plain}
\begin{abstract}
	Communications over power lines in the frequency range above  $2$ MHz, commonly referred to as broadband (BB) power line communications (PLC), has been the focus of increasing research attention and standardization efforts in recent years.  BB-PLC channels are characterized by a dominant colored non-Gaussian additive noise, as well as by periodic variations of the channel impulse response and the noise statistics. In this work we study the fundamental rate limits for BB-PLC channels by bounding  their capacity while accounting for the unique properties of these channels. We obtain explicit expressions for the derived bounds for several BB-PLC noise models, and illustrate the resulting fundamental limits in a numerical analysis.
	
\end{abstract}

\vspace{-0.2cm}
\section{Introduction}
\label{sec:Intro}
\vspace{-0.1cm}
\ac{plc} utilizes the existing power grid infrastructure for data transmission.
Communications over power lines in the frequency range of $2-100$ MHz and possibly beyond, commonly referred to as \ac{bb} \ac{plc} \cite{Ferreira:10}, has received a significant research attention which has supported the development of new standards aiming at facilitating communications at higher data rates \cite{Cano:16}.
Since the indoor power line physical infrastructure consists of three wires,   it is possible to utilize  multiple input ports and/or multiple output ports at terminals by transmitting and/or receiving over multiple differential wire pairs \cite{Berger:15}, thereby realizing \ac{mimo} communications over \ac{bb}-\ac{plc} channels.
The increasing importance of \ac{bb}-\ac{plc} as a high-speed communications medium constitutes a strong motivation for characterizing  the fundamental rate limits of these channels and the associated optimal channel coding schemes.

A major challenge in characterizing the capacity of \ac{bb}-\ac{plc} channels, both for scalar and for \ac{mimo} scenarios, follows since the additive noise in \ac{bb}-\ac{plc} systems is a superposition of several noise sources, including stationary noise, non-impulsive noise with periodic statistics, impulsive noise with periodic statistics, and impulsive noise with non-periodic statistics \cite[Ch. 2.6]{Ferreira:10}.
The resulting overall \ac{bb}-\ac{plc} noise is generally modeled as a non-Gaussian \cite{Meng:05, Gianaroli:14, Mathur:14,  Mathur:15, Bo:07, Gotz:04}, temporally correlated \cite{Gotz:04, Gianaroli:14, Tonello:14, Esmailian:03, Galli:11}, non-stationary  \cite{Gotz:04, Corripio:06, Zimmermann:02b, Cortes:10, Ma:05} process, and \ac{mimo} \ac{bb}-\ac{plc} noise components at different output ports are typically assumed to be correlated \cite{Cano:16, Berger:15, Rende:11}.
The \ac{cir} in BB-PLC channels is typically modeled as a multipath channel \cite{Gotz:04, Zimmermann:02a} with periodic variations \cite{Corripio:06, Gianaroli:14a, Corripio:11}, where the channel outputs typically contain crosstalk from other wires \cite{Cano:16, Berger:15, Pagani:16, Corchado:16, Vernosi:11}.
Common models for the marginal \ac{pdf} of \ac{bb}-\ac{plc} noise include the Nakagami-$m$ distribution \cite{Meng:05}, the Middleton class A distribution \cite{Middleton:77}, and the \ac{gm} distribution \cite{Gianaroli:14,Nassar:11}. All these models characterize only the {\em marginal \ac{pdf}} of the additive noise process, while the {\em complete statistics} of the noise process (i.e., the joint \ac{pdf} of any finite set of sample times) has not been characterized.
The temporal correlation of the {\em stationary} noise component  is typically characterized via its \ac{psd}, for which various models have been proposed \cite{Tonello:14, Esmailian:03, Galli:11}.
The statistics of the {\em periodic} noise component in \ac{bb}-\ac{plc} is  commonly modeled as a cyclostationary process, see \cite{Corripio:06, Cortes:10}.
Lastly,  the non-periodic impulsive noise component in \ac{bb}-\ac{plc} was modeled in \cite{Zimmermann:02b, Ma:05} as a non-stationary process, where \cite{Zimmermann:02b} modeled the arrival times of the impulses using a partitioned Markov chain,  while \cite{Ma:05} modeled these arrival times as a Poisson process.


To avoid the technical difficulties that arise when analyzing the capacity of BB-PLC channels using the accurate statistics of the noise, previous works which attempted to characterize the fundamental rate limits for this channel, used very simplified models which do not capture many of the special characteristics of the noise in \ac{bb}-\ac{plc} channels:
The work \cite{Tonello:14} evaluated the capacity of \ac{bb}-\ac{plc} channels modeled as \ac{lti} systems with additive colored stationary Gaussian noise;
the work \cite{Lampe:13} modeled \ac{bb}-\ac{plc} channels as \ac{lptv} channels with \ac{awgn}, and evaluated the achievable rate by using a transmission scheme which utilizes \ac{ofdm} signalling;
the work \cite{Bo:07} modeled the noise of \ac{bb}-\ac{plc} channels as a Middleton class A process and used the expression for the capacity of \ac{lti} channels with colored stationary Gaussian noise (see, e.g., \cite[Eq. (9.97)]{Cover:06}) to evaluate the capacity.
As this expression was derived for a stationary Gaussian noise, then naturally it does not apply to Middleton class A noise.
We emphasize that all the works mentioned above, i.e., \cite{Tonello:14, Lampe:13, Bo:07}, derived expressions assuming {\em Gaussian noise}, while major works have concluded that the noise is non-Gaussian, see, e.g., \cite{Gotz:04, Meng:05, Gianaroli:14}.
We also note the work \cite{Le:16}, which derived an approximate expression for the achievable rates when using Gaussian inputs and when using inputs with discrete amplitudes, for memoryless channels with additive \ac{gm} noise, which were used for modelling communications in the presence of co-channel interference.
Finally, we note that the capacity of \ac{plc} channels in the {\em narrowband} frequency range ($0-500$ kHz), modeled as additive noise channels in which the CIR is modeled as an \ac{lptv} filter and the noise is a cyclostationary Gaussian process, was derived in  \cite{Shlezinger:15}.
To the best of our knowledge, the fundamental limits for \ac{bb}-\ac{plc} channels, accounting for their unique characteristics, including the {\em non-Gaussianity and the temporal correlation of the noise}, as well as {\em the periodic variations of the \ac{cir} and of the noise statistics}, have not been characterized to date. In this work we aim to address this gap.

{\bf {\slshape Main Contributions}:}
In this work we study the fundamental rate limits of \ac{dt} \ac{bb}-\ac{plc} channels.
We consider a general channel model accounting for a wide range of the characteristics of \ac{bb}-\ac{plc} channels, 
in which the \ac{cir} is modeled as an \ac{lptv} filter,  and the additive noise is modeled as a temporally correlated {\em non-Gaussian cyclostationary process}\footnote{Although the cyclostationary noise statistics does not fully capture the statistics of the non-stationary component  of the \ac{bb}-\ac{plc} noise, it is considered an adequate representation of  the overall temporal variations of the statistics of the additive noise in \ac{bb}-\ac{plc}, see, e.g., \cite{Cortes:10} and \cite[Sec. III-F]{Cano:16}.}.
Accordingly, we characterize an upper bound and two lower bounds on the capacity of these channels.
We note that when the noise is not a Gaussian process, obtaining a closed-form expression for the capacity is generally a very difficult task, even for stationary and memoryless channels, and the common approach is to characterize upper and lower bounds on the capacity, see, e.g., \cite[Ch. 7.4]{Gallager:68}.
To facilitate the derivation, we first derive bounds on the capacity of a general \ac{lti} \ac{mimo} channel with additive {\em non-Gaussian stationary noise}. 
\label{txt:Contributions1}
Then, we prove that the capacity of \ac{bb}-\ac{plc} channels can be obtained from the capacity of non-Gaussian \ac{lti} \ac{mimo} channels by properly setting the parameters of the model, and finally we apply the bounds on the capacity of the latter model to obtain the bounds on the capacity of \ac{bb}-\ac{plc} channels. This approach yields capacity bounds which depend on the \ac{pdf} of the noise process {\em only through its entropy rate and autocorrelation function}. Consequently, our bounds can be obtained explicitly whenever the entropy rate and the autocorrelation function of the noise are known, or can be well-approximated.
 Next, we  derive explicit expressions for the entropy rates for several noise models applicable to \ac{bb}-\ac{plc}, and use them to explicitly characterize the capacity bounds. We also identify scenarios corresponding to known \ac{bb}-\ac{plc} channel models, e.g., \cite{Meng:05,  Mathur:14, Mathur:15, Gianaroli:14}, in which the capacity bounds depend only on the {\em marginal \ac{pdf}} and the autocorrelation function of the noise. In such scenarios  the bounds can be explicitly obtained even when the complete statistical characterization of the noise {process} is unknown.
\label{txt:NakagamiInt1}
The proposed capacity bounds hold for any noise model and distribution. As an example  of our results,  we numerically evaluate the capacity for several \ac{bb}-\ac{plc} noise models, including \ac{gm}, Middleton Class A, and the less common Nakagami-$m$ model. 
Our results demonstrate that, in the high \ac{snr} regime,  the achievable rate of {\em cyclostationary Gaussian signaling} is within a small gap of capacity.
\label{txt:CmmtExhaust}
We also clearly show that assuming the noise is Gaussian may result in significantly underestimating the capacity, and eventually, lead to the design of schemes whose achievable rates are considerably lower than the maximal bit rate that can be supported by the channel.


The rest of this paper is organized as follows: Section \ref{sec:Preliminaries} details the problem formulation;
Section \ref{sec:SecCap} derives bounds on the capacity of \ac{bb}--\ac{plc} channels;
Section \ref{sec:SecBBPLC} presents an application of the results to the characterization of the capacity for several common \ac{bb}-\ac{plc} models which previously appeared in the literature;
Section \ref{sec:Simulations} presents numerical examples;
Lastly, Section \ref{sec:Conclusions} provides some concluding remarks.
Detailed proofs of the results are provided in the appendix.

\vspace{-0.25cm}
\section{Problem Definition}
\label{sec:Preliminaries}
\vspace{-0.1cm}
\subsection{Notations}
\label{subsec:Pre_Notations}
\vspace{-0.1cm}
We use upper-case letters, e.g., $X$, to denote \acp{rv},  lower-case letters, e.g., $x$, to denote deterministic values,  and calligraphic letters, e.g., $\mathcal{X}$, to denote sets.
Column vectors are denoted with boldface letters, e.g., ${\bf{x}}$ for  a deterministic vector and ${\bf X}$ for a random vector;  the $i$-th element of ${\bf{x}}$ ($i \geq 0$) is denoted with $({\bf{x}})_i$.
We use  Sans-Sarif fonts to denote matrices, e.g., $\mathsf{A}$,   the element at the $i$-th row and the $j$-th column of $\mathsf{A}$ is denoted with  $(\mathsf{A})_{i,j}$,  the all-zero $k \times l$ matrix is denoted with $\mathsf{0}_{k\times l}$, and  the $n\times n$ identity matrix is denoted with $\myMat{I}_n$.
Complex conjugate, transpose, Hermitian transpose,  Euclidean norm, stochastic expectation, covariance, differential entropy, and mutual information are denoted by $(\cdot)^*$, $(\cdot)^T$, $(\cdot)^H$, $\left\|\cdot\right\|$, $\E\{ \cdot \}$, $\Cov(\cdot)$, $h(\cdot)$,  and $I(\cdot; \cdot)$,  respectively,
and we use
 $a^+$ to denote $\max\left\{0,a\right\}$, and $\left|\cdot\right|$ to denote the magnitude when applied to scalars, and the determinant  when applied to matrices.
The sets of  non-negative integers, integers, and real numbers are denoted by  $\mathcal{N}$, $\mathcal{Z}$, and $\mathcal{R}$, respectively.
All logarithms are taken to base-2.
Lastly, for any sequence, possibly multivariate, ${\bf y}[i]$, $i \in \mathcal{Z}$, and integers $b_1 < b_2$,  ${\bf y}_{b_1}^{b_2}$ denotes the column vector obtained by stacking $\left[\left( {\bf y}[b_1]\right) ^T,\ldots,\left( {\bf y}[b_2]\right) ^T\right]^T$ and ${\bf y}^{b_2} \equiv {\bf y}_{0}^{b_2}$.

\vspace{-0.2cm}
\subsection{Definitions}
\label{subsec:Pre_Defs}
\vspace{-0.1cm}
In the work we make use of the following definitions:
\vspace{-0.1cm}
\begin{definition}[A \ac{mimo} channel with finite-memory]
	\label{def:Channel}
	A \ac{dt} $\nr \times \nt$ \ac{mimo} channel with finite memory
	consists of an input sequence $\Xin[i] \in \mathcal{R}^{\nt}$, $i \in \mathcal{N}$, an output sequence  $\Yi[i] \in \mathcal{R}^{\nr}$, $i \in \mathcal{N}$, an initial state vector ${\bf S}_0 \in \mathcal{S}_0$ of finite dimensions, and a sequence of \acp{pdf} 
	$\big\{p_{\Yi^{n}|\Xin^{n},{\bf S}_0}\left(\yi^{n}|\xin^{n},{\bf s}_0\right)\big\}_{n=0}^{\infty}$.
\end{definition}

\begin{definition}[Code]
	\label{def:Code}
		An $\left[R, l \right]$ code with rate $R$ and blocklength $l \in \mathcal{N}$ consists of:
		{\em 1)} A message set $\mathcal{U} \triangleq \{1,2,\ldots,2^{lR}\}$.
		{\em 2)} An  encoder $e_l$ which maps each message $u \in \mathcal{U}$ into an $\nt \times l$ codeword matrix $\myMat{X}_{(u)}^{l-1} \triangleq  \big[\xin_{(u)}\left[0\right],\xin_{(u)}\left[1\right],\ldots,\xin_{(u)}\left[l-1\right]\big]$, where $\xin_{(u)}\left[i\right]$ denotes the inputs at the $\nt$ channel input ports at time $i$.
		{\em 3)}  A decoder $d_l$ which maps the channel output sequence  $\big[{\bf y} \left[0\right], {\bf y} \left[1\right],\ldots, {\bf y}\left[l-1\right]\big] \in \mathcal{R}^{\nr \times l}$ into a message $\hat{u} \in \mathcal{U}$.
		The encoder and decoder operate independently of the initial state, in the sense that ${\bf S}_0$  does not affect the encoding and the decoding operations.
%
%
\end{definition}
The set $\big\{\myMat{X}_{(u)}^{l-1}\big\}_{u =1}^{2^{lR}}$ is referred to as the {\em codebook} of the $\left[R, l \right]$ code.
Assuming the message $U$ is uniformly selected from $\mathcal{U}$, the average probability of error, when the initial state is ${\bf s}_0$, is:
		\vspace{-0.1cm}
\begin{equation*}
\label{eqn:def_avgError}
P_{\rm e}^l \left( {{\bf s} }_0 \right) \!= \! \frac{1}{2^{lR}}\sum\limits_{u \!= \! 1}^{2^{lR}} \Pr \left( {\left. {{d_l }\left( \Yi^{l\!- \!1}  \right) \ne u} \right|U \!= \! u,{{{\bf S} }_0} \!= \! {{{\bf s} }_0}} \right).
		\vspace{-0.2cm}
\end{equation*}
%
\begin{definition}[Achievable rate]
	\label{def:Rate}
	A rate $\Rs$ is called achievable if, for every $\epsilon_1,\epsilon_2 > 0$, there exists a positive integer $l_0 >0$ such that for all integer $l > l_0$, there exists an $\left[R, l \right]$ code which satisfies 
	$\mathop {\sup }\limits_{{{{\bf s} }_0} \in \mathcal{S}_0} P_{\rm e}^l \left( {{{{\bf s} }_0}} \right) < {\epsilon _1}$, and $R \geq \Rs - \epsilon_2$.
\end{definition}

\begin{definition}[Capacity]
		\label{def:Capacity}
	{Capacity} is defined as the supremum of all achievable rates.	
\end{definition}
%
\vspace{-0.3cm}
\subsection{Model and Problem Formulation}
\label{subsec:Pre_Model}
\vspace{-0.1cm}
We consider a \ac{dt} $\tnr \times \tnt$ \ac{mimo} \ac{bb}-\ac{plc} channel with $\tnr$ receive ports and $\tnt$ transmit ports, modeled as a multivariate \ac{lptv} system with additive non-Gaussian cyclostationary noise\footnote{\label{ftn:Tilde}In the following, we use the tilde notation for quantities associated with the   \ac{mimo} \ac{bb}-\ac{plc} channel, highlighting the fact that this is a periodic channel model. The same notations without a tilde represent the corresponding quantities associated with the  \acl{lgmc} defined in Subsection \ref{subsec:LGMC}, which is a non-periodic channel model.
}.
Let $\Memeq$ be a non-negative integer which represents the {\em length of the memory of the channel}, $\PerCh$ be a positive integer which represents the {\em period of the CIR}, and $\PerN$ be a positive integer which represents the {\em period of the noise statistics}.
Let $\Weq[i] \in \mathcal{R}^{\tnr}$ be a real-valued $\tnr$-dimensional zero-mean strict-sense cyclostationary non-Gaussian additive noise\footnote{\label{ftn:Cyclo}
Previous works which studied the cyclostationarity of \ac{bb}-\ac{plc} noise, \cite{Corripio:06, Cortes:10},  did not explicitly conclude whether the noise process is cyclostationary in the  strict-sense  or in the  wide-sense. We note that  in \cite[Sec. III-F]{Cortes:10} it is observed that the marginal \ac{pdf} of the noise is periodic, which  is an indication that the noise process can be modeled as a strict-sense cyclostationary process.}, i.e., for any set of $k$ integer indexes $\{i_l\}_{l=1}^k$, $k\in\mN$, the joint \ac{pdf} of $\Weq[i_1],\Weq[i_2],\ldots,\Weq[i_k]$ is equal to the joint \ac{pdf} of $\Weq[i_1+\PerN],\Weq[i_2+\PerN],\ldots,\Weq[i_k+\PerN]$.
Since the channel memory is $\Memeq$, then noise vectors more than $\Memeq$ instances apart are mutually independent,  i.e., $\forall i_1,i_2,l_1,l_2 \in \mathcal{N}$ such that $i_2 > i_1 + l_1 + \Memeq$, the random vectors $\Weq_{i_1}^{i_1+l_1}$ and $\Weq_{i_2}^{i_2+l_2}$ are mutually independent.
We further assume that there is no deterministic dependence between instances of $\Weq[i]$, i.e., $\nexists i_0$ for which  $\Weq[i_0]$  can be expressed as a linear combination of $\big\{\Weq[i]\big\}_{i \neq i_0}$.
Let $\big\{ \Heq [i, \tau ] \big\}_{\tau  = 0}^{\Memeq}$ denote the \ac{lptv} \ac{cir}. The periodicity of the \ac{cir} implies that  $\Heq [i, \tau ] = \Heq [i + \PerCh, \tau ]$, $\forall i \in \mathcal{Z}, \tau \in \{0,1,\ldots,\Memeq\}$.

With the above definitions, the input-output relationship for the \ac{mimo} \ac{bb}-\ac{plc} channel with input codeword length $\tilde{l}$ is given by
\vspace{-0.2cm}
\begin{equation}
	\label{eqn:RxModel_1}
	\Yeq[i] = \sum\limits_{\tau = 0}^{\Memeq} \Heq[i, \tau]\Xeq[i - \tau]  + \Weq[i], \quad i \in \{0,1,\ldots,\tilde{l} - 1\},
	\vspace{-0.1cm}
\end{equation}
where the initial state of the channel (i.e., prior to the beginning of reception) is given by ${\tilde{\bf S} _0} = \left[ \big(\Xeq _{ - \Memeq}^{-1}\big)^T,\big(\Weq _{ - \Memeq}^{-1}\big)^T\right]^T$.
The channel input is subject to a time-averaged power constraint $\PCsteq$, as in \cite[Eq. (7)]{Shlezinger:15} and \cite[Eq. (7)]{Goldsmith:01}:
\vspace{-0.2cm}
\begin{equation}
\label{eqn:Constraint0}
\frac{1}{\tilde{l}}\sum\limits_{i=0}^{\tilde{l}-1}\E \left\{\left\| \Xeq\left[ i \right] \right\|^2\right\} \le \PCsteq.
\vspace{-0.1cm}
\end{equation}

Set $\Per$ to be the least common multiple\footnote{\label{ftn:SyncSampl} The common practice in BB-PLC systems, namely, sampling at a rate which is an integer multiple of twice the AC frequency, typically results in $\PerCh = \PerN$  or $\PerCh = 2\PerN$ \cite{Corripio:06}. In this work we allow a general  relationship between the periods of the \ac{cir} and of the noise statistics, but still assume synchronized sampling, i.e., we assume that the sampling period is a rational multiple of the period of the continuous-time \ac{cir} as well as of the period of the statistics of the continuous-time  noise signal. Allowing a general relationship facilitates accommodating additional \ac{bb}-\ac{plc} scenarios, e.g., interference-limited \ac{bb}-\ac{plc}, by our framework.} of $\PerCh$ and $\PerN$ which satisfies $\Per > \Memeq$.
As the \ac{cir} and the statistics of the noise of the \ac{bb}-\ac{plc} channel \eqref{eqn:RxModel_1} are both periodic with period $\Per$, we  refer to $\Per$ as the {\em period of the channel}.
\label{txt:ComplexRep}
While the above model was stated for real signals, complex (baseband) \ac{bb}-\ac{plc} channels can be  accommodated by this model by representing all complex vectors and matrices using real vectors and matrices having twice - for vectors, and four times - for matrices, the number of elements, corresponding to the real and to the imaginary parts of the complex components, see, e.g., \cite[Sec. I]{Weingerten:06}. Accordingly, a complex \ac{mimo} \ac{bb}-\ac{plc} channel with an $\tnt\Complex \times 1$ complex input $\Xeq\Complex[i]$, an $\tnr\Complex \times 1$ complex output $\Yeq\Complex[i]$, an $\tnr\Complex \times 1$ complex additive noise $\Weq\Complex[i]$, and an $\tnr\Complex \times \tnt\Complex$ \ac{cir} $\big\{ \Heq\Complex [i, \tau ] \big\}_{\tau  = 0}^{\Memeq}$, can be equivalently represented as a real \ac{mimo} \ac{bb}-\ac{plc} channel corresponding to \eqref{eqn:RxModel_1}, via the statement in \eqref{eqn:ComplexRep1}. 
	\begin{equation}
	\label{eqn:ComplexRep1}
	\left[ {\begin{array}{*{20}{c}}
		{\rm Re}\Big\{ \Yeq\Complex[i]  \Big\} \\
		{\rm Im}\Big\{  \Yeq\Complex[i]  \Big\}
		\end{array}} \right]
	= \sum\limits_{\tau = 0}^{\Memeq} 
	\left[ {\begin{array}{*{20}{c}}
		{\rm Re}\Big\{  \Heq\Complex [i, \tau ]  \Big\} &\hspace{0.2cm} -{\rm Im}\Big\{ \Heq\Complex [i, \tau ]\Big\} \\
		{\rm Im}\Big\{  \Heq\Complex [i, \tau ]  \Big\} &\hspace{0.2cm} {\rm Re}\Big\{ \Heq\Complex [i, \tau ]\Big\}
		\end{array}} \right]
	\left[ {\begin{array}{*{20}{c}}
		{\rm Re}\Big\{  \Xeq\Complex[i - \tau]  \Big\} \\
		{\rm Im}\Big\{  \Xeq\Complex[i - \tau]  \Big\}
		\end{array}} \right]
	+ 	\left[ {\begin{array}{*{20}{c}}
		{\rm Re}\Big\{  \Weq\Complex[i]  \Big\} \\
		{\rm Im}\Big\{  \Weq\Complex[i]  \Big\}
		\end{array}} \right].
	\end{equation}

In the following section we study the capacity of the \ac{mimo} \ac{bb}-\ac{plc} channel defined above subject to a time-averaged power constraint $\PCsteq$. The capacity of this channel is denoted as $\Capeq$.

\vspace{-0.2cm}
\section{The Capacity of \ac{mimo} \ac{bb}-\ac{plc} Channels} 
\label{sec:SecCap}
\vspace{-0.1cm}
Our main result is the characterization of upper and lower bounds on the capacity of  \ac{mimo} \ac{bb}-\ac{plc} channels, defined in \eqref{eqn:RxModel_1}. This result is obtained via the following three steps:
\begin{itemize}
	\item First, in Subsection \ref{subsec:LGMC}, we define a general \ac{lti}  $\nr \times \nt$  \ac{mimo} channel with stationary non-Gaussian noise, to which we refer as the {\em \ac{lgmc}}.
	We express the capacity of the LNGMC as a limit of the maximum mutual information between its	input and its output as the blocklength increases to infinity.
	\item Next, we derive computable upper and lower bounds on the capacity of the \ac{lgmc}, which are stated in terms of the \ac{cir}, and of the entropy rate and autocorrelation of the  noise.
	\item Lastly, in Subsection \ref{subsec:BBPLC}, we prove that the capacity of the \ac{bb}-\ac{plc} channel can be obtained as the capacity of an equivalent $\Per \times \Per$ \ac{lgmc}, and use the bounds obtained to state the corresponding capacity bounds for the \ac{bb}-\ac{plc} channel.
\end{itemize}

\vspace{-0.2cm}
\subsection{Analysis of the Capacity of the \ac{lgmc}}
\label{subsec:LGMC}
\vspace{-0.1cm}
We begin with the definition of the \ac{lgmc}:
Let $\Mem$ be a non-negative integer which represents the {\em length of the memory of the channel}, and let $\left\{ \Hi [\tau ] \right\}_{\tau  = 0}^{\Mem}$ denote a set of $\Mem+1$ real-valued $\nr \times \nt$ \ac{lti} channel transfer matrices.
Let $\Wi[i] \in \mathcal{R}^{\nr}$ be a multivariate, real-valued, strict-sense stationary non-Gaussian additive noise process, whose mean is zero and whose temporal dependence spans a finite interval of length $\Mem$, i.e., $\forall i_1,i_2,l_1,l_2 \in \mathcal{N}$ such that $i_2 > i_1 + l_1 + \Mem$, the random vectors $\Wi_{i_1}^{i_1+l_1}$ and $\Wi_{i_2}^{i_2+l_2}$ are mutually independent.
For the transmission of a block of $l$ symbols, the input-output relationship for the channel is given by
\vspace{-0.2cm}
\begin{equation}
\label{eqn:RxModel_2}
\Yi[i] = \sum\limits_{\tau = 0}^{m} \Hi[\tau]\Xin[i - \tau]  + \Wi[i], \quad i \in \{0,1,\ldots,l - 1\},
\vspace{-0.1cm}
\end{equation}
where the initial state of the channel is given by ${{\bf S} _0} = \left[\big(\Xin _{ - m}^{-1}\big)^T, \big(\Wi _{ - m}^{-1}\big)^T \right]^T$.
The channel input is subject to a time-averaged power constraint $\PCst$, i.e.,
\vspace{-0.2cm}
\begin{equation}
\label{eqn:Constraint1}
\frac{1}{l}\sum\limits_{i=0}^{l-1}\E \left\{\left\| \Xin\left[ i \right] \right\|^2\right\} \le \PCst.
\vspace{-0.1cm}
\end{equation}
\label{txt:SpecialCase}
While the definition of the \ac{lgmc} in \eqref{eqn:RxModel_2}--\eqref{eqn:Constraint1} can be obtained as a special case of the definition of the \ac{mimo} \ac{bb}-\ac{plc} channel in \eqref{eqn:RxModel_1}--\eqref{eqn:Constraint0} by setting the period to unity, we use Eqs. \eqref{eqn:RxModel_2}--\eqref{eqn:Constraint1} to highlight the fact that the \ac{lgmc} is non-periodic and to introduce the different quantities associated with the model separately from the periodic \ac{mimo} \ac{bb}-\ac{plc} channel model.

The capacity of the \ac{lgmc} defined above is stated in the following proposition:
\begin{proposition}
	\label{pro:AsymCapEq}
	The capacity of the \ac{lgmc} defined in \eqref{eqn:RxModel_2} subject to  \eqref{eqn:Constraint1} is given by
	\vspace{-0.1cm}
	\begin{equation}
	\label{eqn:AsymCapEq}
	\Capacity = \mathop{\lim}\limits_{n \rightarrow \infty} \frac{1}{n}\mathop{\sup}\limits_{\Pdf{\Xin^{n-1}}: \frac{1}{n}\sum\limits_{i=0}^{n-1}\E \left\{\left\| \Xin\left[ i \right] \right\|^2\right\} \le \PCst  } I\left(\Xin^{n-1} ; \Yi^{n-1} | \Xin_{-\Mem}^{-1} = {\bf 0}_{\nt \cdot \Mem} \right).
	\vspace{-0.1cm}
	\end{equation}
%
\end{proposition}
\begin{IEEEproof}
 Note that \eqref{eqn:AsymCapEq} corresponds to the capacity of an {\em information stable} channel \cite{Dobrushin:63}. Information stable channels can be roughly described as having the property that the input which maximizes the mutual information and its corresponding output behave	ergodically, thus information stability depends on the conditional \ac{pdf} of the channel output given the channel input.
  Since stationary channels with finite memory are known to be information stable\footnote{
  The information stability of stationary channels with finite memory, in which the input and the output are taken from {\em discrete and finite alphabets}, was shown in \cite{Tsaregradskii:58}, see also \cite[Sec. 1.5]{Dobrushin:63}. This results also holds for arbitrary alphabets, see  \cite[Thm. 6]{Han:03}.}, see, e.g., \cite[Sec. 1.5]{Dobrushin:63}, the proposition follows.
\end{IEEEproof}

%
%
\vspace{-0.1cm}
\begin{remark}
	\label{rem:Pro1_0}
	{\em Previous works on the capacity of finite-memory channels with Gaussian noise, e.g., \cite{Massey:88,   Goldsmith:01}, obtained a capacity result in the frequency domain, by transforming the  channel into a set of parallel independent channels, which allows expressing capacity as an explicit integral. When the noise is non-Gaussian, switching to the frequency domain still results in the noise components at different frequency bins having statistical dependence (even if the noise samples are independent in the time domain), and consequently switching to the frequency-domain in such cases will typically not yield  a set of parallel independent channels. For this reason, our analysis is carried out in the time domain, and the capacity has to be stated in terms of an asymptotic limit. 
	Nonetheless, the {\em bounds} on the capacity of \acp{lgmc},  derived in Props. \ref{pro:UpperBoundCap} and \ref{pro:LowerBoundCap}, are stated explicitly (not as limit expressions) in the frequency domain.	
	}
\end{remark}
\vspace{-0.1cm}
 Prop. \ref{pro:AsymCapEq} implies that the capacity of the \ac{lgmc} can be computed by setting $ \Xin_{-\Mem}^{-1} = {\bf 0}_{\nt \cdot \Mem}$. We note that setting the signal component in the initial state to zero was used as a model assumption in \cite{Massey:88} and \cite{Brandenburg:74}, which studied the capacity of point-to-point channels with memory and Gaussian noise.
 Note that by defining the $l\cdot \nr \times l \cdot \nt$ matrix $\tilde{\Hi}_l$ such that
		\vspace{-0.1cm}
		\begin{equation}
		\label{eqn:DefHTilde}
		\tilde{\Hi}_l \!\triangleq\! \left[ {\begin{array}{*{20}{c}}
			{\Hi\!\left[0\right]}&\cdots &0&  \cdots &0\\
			\vdots & \ddots &{}&\ddots& \vdots \\
			{\Hi\!\left[\Mem\right]}& \cdots &{\Hi\!\left[0\right]}& \cdots &0\\
			\vdots & \ddots &{}& \ddots & \vdots\\
			0& \cdots &{\Hi\!\left[\Mem\right]}& \cdots &{\Hi\!\left[0\right]}
			\end{array}} \right],
		\vspace{-0.1cm}
		\end{equation}
	 and setting $ \Xin_{-\Mem}^{-1} = {\bf 0}_{\nt \cdot \Mem}$, the received signal samples can be expressed as
	 		\vspace{-0.1cm}
		\begin{equation}
		\label{eqn:EqRelation}
		\Yi^{l-1} = \tilde{\Hi}_l\Xin^{l-1} + \Wi^{l-1}.
		\vspace{-0.1cm}
		\end{equation}

Next, based on the capacity expression in Prop. \ref{pro:AsymCapEq}, we derive upper and lower bounds on $\Capacity$, which depend on the distribution of the non-Gaussian noise $\Wi[i]$ only through its autocorrelation function, $\CWi[\tau] \triangleq \E\left\{\Wi[i + \tau]\big( \Wi[i]\big)^T \right\}$, and its entropy rate, $\HWi \triangleq \mathop{\lim}\limits_{l\rightarrow \infty}\frac{1}{l}h\left(\Wi^{l-1}\right)$. Note that the strict-sense stationarity and finite temporal dependence of $\Wi[i]$ imply that the entropy rate limit exists and that it equals  $\HWi  = h\left(\left.\Wi[\Mem]\right| \Wi^{\Mem-1} \right)$ \cite[Ch. 12.5]{Cover:06}.

In the statement of the bounds we make use of the following additional definitions: For any $\omega\! \in\! \left[-\pi,\pi\right)$, we define the $\nr \!\times\!  \nt$ matrix $\Hfd(\omega) \!\triangleq \! \sum\limits_{\tau=0}^\Mem\!\Hi[\tau]e^{-j\omega \tau}$, and
the  $\nr\! \times\! \nr$ matrix $\CWfd(\omega)\! \triangleq\!  \sum\limits_{\tau=-\Mem}^\Mem\CWi[\tau]e^{-j\omega \tau}$,
and we let $\{\TlambdaN _k (\omega)\}_{k=0}^{\nr-1}$ and $\{\Tlambda _k (\omega)\}_{k=0}^{\nr-1}$ denote the eigenvalues of $\Hfd(\omega)\big(\Hfd(\omega)\big)^H$ and of $\big(\Hfd(\omega)\big)^H\big(\CWfd(\omega)\big)^{-1}\Hfd(\omega)$, respectively.
Next, let $\HWiG$ be the entropy rate of a zero-mean $\nr \times 1$ multivariate {\em Gaussian} process with autocorrelation function $\CWi[\tau]$, and let $\CapGauss$ be the capacity of the channel defined in \eqref{eqn:RxModel_2} subject to the constraint \eqref{eqn:Constraint1} and to the setting $\Xin_{-\Mem}^{-1} = {\bf 0}_{\nt \cdot \Mem}$, when the noise $\Wi[i]$ is {\em Gaussian}. From \cite[Sec. III]{Verdu:88} the entropy rate $\HWiG$ can be expressed as
\begin{subequations}
	\label{eqn:IntgDefs}
	\vspace{-0.2cm}
	\begin{equation}
		\label{eqn:IntgDefs1}
		\HWiG = \frac{1}{4\pi }\int\limits_{\omega = -\pi}^{\pi}\log \left|2 \pi e \CWfd(\omega)   \right|d \omega.
	\vspace{-0.1cm}
	\end{equation}
	In 	\cite[Eqn. (9)]{Brandenburg:74} the capacity of \ac{mimo} channels with an \ac{lti} \ac{cir} and additive stationary Gaussian noise was characterized\footnote{We note that  \cite[Thm. 1]{Brandenburg:74}  is stated for a per-codeword power constraint. However, it follows from \cite[Sec. 3.1]{Brandenburg:74} and from \cite[Ch. 7.3, pgs. 323-324]{Gallager:68} that the proof of \cite[Thm. 1]{Brandenburg:74} also holds subject to the time-averaged power constraint \eqref{eqn:Constraint1}.}, assuming the signal component in the  initial state is zero (i.e., $\Xin_{-\Mem}^{-1} = {\bf 0}_{\nt \cdot \Mem}$). Using \cite[Eqn. (9)]{Brandenburg:74} we can write the capacity of the
	channel \eqref{eqn:RxModel_2} when $\Wi[i]$ is replaced by a zero-mean stationary Gaussian process  with the same autocorrelation, as
\vspace{-0.1cm}
 	\begin{equation}
 	\label{eqn:IntgDefs2}
 \CapGauss = \frac{1}{4\pi}\sum\limits_{k=0}^{ \nr -1} \int\limits_{\omega = -\pi}^{\pi} \Big(\log \big(\TDelta \cdot {\Tlambda _k (\omega)}\big)\Big)^+ d\omega,
 	\vspace{-0.1cm}
 	\end{equation}
 where $\TDelta$ is selected to satisfy $\frac{1}{2\pi }\sum\limits_{k=0}^{ \nr -1} \int\limits_{\omega = -\pi}^{\pi} \Big(\TDelta - \left(\Tlambda _k (\omega)\right)^{-1}\Big)^+ d\omega =  \PCst$.

 \end{subequations}
 
 \label{txt:Quantities}
Note that $\HWiG$ and $\CapGauss$, defined in \eqref{eqn:IntgDefs}, correspond to the entropy rate of a Gaussian noise process, and to the capacity of a channel with additive Gaussian noise, respectively. These quantities are used for facilitating the characterization of the bounds on the capacity of the \ac{lgmc} in which the noise  is a non-Gaussian process.

We next state an upper bound and two lower bounds on the capacity of the \ac{lgmc} using $\HWi$, $\HWiG$, and $\CapGauss$.
First, the upper bound is stated in the following proposition:
\begin{proposition}
	\label{pro:UpperBoundCap}
	The capacity of the \ac{lgmc} defined in \eqref{eqn:RxModel_2}, subject to the constraint  \eqref{eqn:Constraint1},  satisfies
\vspace{-0.1cm}
	\begin{equation}
	\label{eqn:UpperBoundCap1}
	\Capacity \leq \CapGauss + \HWiG -  \HWi.
\vspace{-0.1cm}	
	\end{equation}

		\noindent
		[A proof is given in Appendix \ref{app:Proof3}]	
\end{proposition}
Next, two lower bounds on the capacity of the \ac{lgmc} are stated in the following Prop. \ref{pro:LowerBoundCap}:
\begin{proposition}
	\label{pro:LowerBoundCap}
	The capacity of the \ac{lgmc} defined in \eqref{eqn:RxModel_2} subject to the constraint  \eqref{eqn:Constraint1}  satisfies
	\begin{subequations}
		\label{eqn:LowerBoundCap1}	
\vspace{-0.2cm}			
		\begin{equation}
		\label{eqn:LowerBoundCap1a}
		\Capacity \geq  \CapGauss.
\vspace{-0.1cm}		
		\end{equation}		
	Moreover, if $\nr = \nt$ and $\Hi[0]$ is invertible, then $\Capacity$ also satisfies 	
\vspace{-0.2cm}	
		\begin{equation}
		\label{eqn:LowerBoundCap1b}
		\Capacity \geq  \frac{\nr}{2} \log\left(\frac{2\pi e \PCst}{\nt} \cdot 2^{\frac{1}{2\pi \cdot \nt}\sum\limits_{k=0}^{ \nr -1} \int\limits_{\omega = -\pi}^{\pi} \log\left( \TlambdaN _k (\omega)\right)  d\omega} + 2^{\frac{2}{\nr}\HWi}\right) - \HWi.
\vspace{-0.1cm}		
		\end{equation}			
	\end{subequations}
	
	\noindent
	[A proof is given in Appendix \ref{app:Proof4}]		
\end{proposition}
%

\vspace{-0.4cm}
\subsection{Capacity Analysis for \ac{mimo} \ac{bb}-\ac{plc} Channels}
\label{subsec:BBPLC}
\vspace{-0.1cm}
In order to obtain bounds on the capacity of  \ac{mimo}  \ac{bb}-\ac{plc} channels, we first prove that any $\tnr \times \tnt$  \ac{mimo}  \ac{bb}-\ac{plc} channel, in which the \ac{cir} and the noise statistics are periodic with a period of $\Per$, can be equivalently represented (in terms of the achievable rates) as an $\Per\cdot\tnr \times \Per\cdot\tnt$ \ac{lgmc}, in which the \ac{cir} is time-invariant and the noise is stationary. Then, we apply the capacity bounds derived for the \ac{lgmc} to bound the capacity of the original \ac{mimo}  \ac{bb}-\ac{plc} channel by considering its equivalent \ac{lgmc} with the appropriate dimensions.
	To that aim,  define two $\Per \cdot \tnr  \times \Per \cdot \tnt$ matrices, $\Hi\BB[0]$ and $\Hi\BB[1]$, as follows:
\vspace{-0.2cm}
	\begin{equation*}
	\Hi\BB[0] \!\triangleq\! \left[ {\begin{array}{*{20}{c}}
		{\Heq\!\left[0,0\right]}&\cdots &0&  \cdots &0\\
		\vdots & \ddots &{}&\ddots& \vdots \\
		{\Heq\!\left[\Memeq,\Memeq\right]}& \cdots &{\Heq\!\left[\Memeq,0\right]}& \cdots &0\\
		\vdots & \ddots &{}& \ddots & \vdots\\
		0& \cdots &{\Heq\!\left[\Per\! - \!1,\Memeq\right]}& \cdots &{\Heq\!\left[\Per\! - \!1,0\right]}
		\end{array}} \right], \hspace{0.1cm}
	\Hi\BB[1] \!\triangleq\! \left[ {\begin{array}{*{20}{c}}
		0& \cdots &0&{\Heq\!\left[0,\Memeq\right]}& \cdots &{\Heq\!\left[0,1\right]}\\
		\vdots &{}& \vdots &{}& \ddots & \vdots \\
		0& \cdots &0&0&{}&{\Heq\!\left[\Memeq\! - \!1,\Memeq\right]}\\
		\vdots &{}& \vdots & \vdots &{}& \vdots \\
		0& \cdots &0&0& \cdots &0
		\end{array}} \right],
	\end{equation*}
and also define the $\Per \cdot \tnr  \times 1$  random vector
$\Wi\BB\left[\,\ieq\,\right] \triangleq \Weq_{\ieq \cdot \Per}^{\left(\ieq \! + \! 1\right)\cdot \Per - 1}$.
As $\Wi\BB\left[\,\ieq\,\right]$ is given by the \ac{dcd} \cite{Giannakis:98} of $\Weq[i]$,  the strict-sense cyclostationarity of $\Weq[i]$ induces a strict-sense stationarity of $\Wi\BB\left[\,\ieq\,\right]$.
Using these definitions, we construct an \ac{lgmc} with a $\Per \cdot \tnt \times 1$ input $\Xin\BB\left[\,\ieq\,\right]$ and a $\Per \cdot \tnr \times 1$ output $\Yi\BB\left[\,\ieq\,\right]$ which satisfies the following input-output relationship for a sequence of $l$ channel inputs:
\vspace{-0.2cm}
\begin{equation}
\label{eqn:RxModel_3}
\Yi\BB\left[\,\ieq\,\right] = \sum\limits_{\teq = 0}^{1} \Hi\BB\left[\teq\right]\Xin\BB\left[\ieq - \teq\right]  + \Wi\BB\left[\,\ieq\,\right], \qquad \ieq \in \{0,1,\ldots,l - 1\},
\vspace{-0.1cm}
\end{equation}
  where the channel input to the \ac{lgmc} \eqref{eqn:RxModel_3} has to satisfy an average power constraint
  \vspace{-0.2cm}
  \begin{equation}
  \label{eqn:ConstraintBB}
  \frac{1}{l}\sum\limits_{\ieq=0}^{l-1}\E \left\{\left\| \Xin\BB\left[\, \ieq\, \right] \right\|^2\right\} \le \PCst\BB =\Per \cdot \PCsteq.
  \vspace{-0.1cm}
  \end{equation}
  Since  $\Per > \Memeq$, the initial state of the \ac{lgmc} is  ${\bf S}_{0, {\rm DCD}} = \big[\Xin\BB^T[-1], \Wi\BB^T[-1]\big]^T$.
   Let $\TCapacity$ be the capacity of the \ac{lgmc} defined in \eqref{eqn:RxModel_3}--\eqref{eqn:ConstraintBB}. The relationship between the capacity of the \ac{bb}-\ac{plc} channel in \eqref{eqn:RxModel_1}--\eqref{eqn:Constraint0} and the  \ac{lgmc} in \eqref{eqn:RxModel_3}--\eqref{eqn:ConstraintBB} is stated in the following theorem:
\begin{theorem}
	\label{thm:MainResult1}
	The capacity of the \ac{bb}-\ac{plc} channel, defined in \eqref{eqn:RxModel_1}, subject to   \eqref{eqn:Constraint0} satisfies
	  \vspace{-0.1cm}
	\begin{equation}
	\label{eqn:MainResult1}
	\Capeq= \frac{1}{\Per} \TCapacity.
	  \vspace{-0.1cm}
	\end{equation}
	
	\noindent
	[A proof is given in Appendix \ref{app:Proof2}]	
\end{theorem}

Based on Thm. \ref{thm:MainResult1} and Props. \ref{pro:UpperBoundCap} and \ref{pro:LowerBoundCap}, we obtain lower and upper bounds on the capacity of the \ac{bb}-\ac{plc} channel.
To that aim, define the $\Per \cdot \tnr\times \Per\cdot \tnr$ autocorrelation function
$\CWiBB[\teq] \triangleq \E\left\{\Wi\BB\left[\ieq + \teq\right]  \Wi\BB^T\left[\,\ieq \,\right]  \right\}$,
the entropy rate
$\HWiBB \triangleq \mathop{\lim}\limits_{n\rightarrow \infty}\frac{1}{n}h\left(\Wi\BB^{n-1}\right)$,
the $\Per\cdot \tnr  \times \Per \cdot \tnt $ matrix
$\Hfd\BB(\omega) \!\triangleq \! \sum\limits_{\teq=0}^1\!\Hi\BB[\teq]e^{-j\omega \teq}$, and
the  $\Per \cdot \tnr \times \Per\cdot \tnr $ matrix
$\CWfdBB(\omega)\! \triangleq\!  \sum\limits_{\teq=-1}^1\CWiBB[\teq]e^{-j\omega \teq}$.
Next, let
$\big\{\TlambdaN _{{\rm DCD},k} (\omega)\big\}_{k=0}^{\Per\cdot \tnr-1}$ and
$\big\{\Tlambda _{{\rm DCD},k} (\omega)\big\}_{k=0}^{\Per\cdot \tnr-1}$ be the eigenvalues of
$\Hfd\BB(\omega)\left(\Hfd\BB(\omega)\right)^H$ and of $\big(\Hfd\BB(\omega)\big)^H\big(\CWfdBB(\omega)\big)^{-1}\Hfd\BB(\omega)$, respectively, and, in addition,
let $\HWiGBB$ denote the entropy rate of a zero mean $\Per \cdot \tnr \times 1$ {\em Gaussian} process with autocorrelation function $\CWiBB[\teq]$. $\HWiGBB$  can be computed via \eqref{eqn:IntgDefs1} with $\CWfdBB(\omega)$ instead of $\CWfd(\omega)$.
Finally, let  $\CapGaussBB$ be the capacity of the \ac{lgmc} \eqref{eqn:RxModel_3} when the noise $\Wi\BB\left[\,\ieq\,\right]$ is {\em Gaussian} with autocorrelation function $\CWiBB[\teq]$. Thus, $\CapGaussBB$ is obtained using \eqref{eqn:IntgDefs2} with $\Tlambda _{{\rm DCD},k} (\omega)$ and $\PCst\BB$ replacing $\Tlambda _{k} (\omega)$ and $\PCst$, respectively.
Noting that $\Hi\BB[0]$ has a full rank if and only if $\Heq\left[\ieq,0\right]$ has a full rank for every $\ieq \in \{0,1,\ldots,\Per-1 \}\triangleq \tilde{\PerSet}$ \cite[Ex. 3.7.4]{Meyer:00}, then, by combining Thm. \ref{thm:MainResult1} with  Prop. \ref{pro:UpperBoundCap}, the following upper bound on $\Capeq$ is obtained:
\begin{corollary}
	\label{cor:UpperBoundBB}
	The capacity of the \ac{bb}-\ac{plc} channel defined in \eqref{eqn:RxModel_1}, subject to \eqref{eqn:Constraint0}, satisfies
		\begin{equation}
		\label{eqn:UpperBoundBB1}
		\Capeq \leq \frac{1}{\Per}\left( \CapGaussBB + \HWiGBB -  \HWiBB\right) .
		\end{equation}
\end{corollary}
Lastly, combining Thm. \ref{thm:MainResult1} with  Prop. \ref{pro:LowerBoundCap}, the following lower bounds on $\Capeq$ are obtained:
\begin{corollary}
	\label{cor:LowerBoundBB}
	The capacity of the \ac{bb}-\ac{plc} channel defined in \eqref{eqn:RxModel_1}, subject to  \eqref{eqn:Constraint0}, satisfies
	\begin{subequations}
		\label{eqn:LowerBoundCBB1}		
		\begin{equation}
		\label{eqn:LowerBoundBB1a}
		\Capeq \geq  \frac{1}{\Per}\CapGaussBB.
		\end{equation}		
		Moreover, if   $\tnr = \tnt$ and $\Heq\left[\,\ieq,0\right]$ is non-singular for every $\ieq \in \tilde{\PerSet}$, then $\Capeq$ also satisfies 	
		\begin{equation}
		\label{eqn:LowerBoundBB1b}
		\Capeq \geq  \frac{\tnt}{2} \log\bigg(\frac{2\pi e \PCsteq}{\tnt} \cdot 2^{\frac{1}{2\pi \cdot \Per \cdot \tnt}\sum\limits_{k=0}^{ \Per \cdot \tnr -1} \int\limits_{\omega = -\pi}^{\pi} \log\left( \TlambdaN _{{\rm DCD},k} (\omega)\right)  d\omega} + 2^{\frac{2}{\Per \cdot \tnr}\HWiBB}\bigg) - \frac{1}{\Per}\HWiBB.
		\end{equation}			
	\end{subequations}	
\end{corollary}
\vspace{-0.1cm}
\begin{remark}
	\label{rem:CorBB_1}
	{\em From the proof of Prop. \ref{pro:LowerBoundCap} in Appendix \ref{app:Proof4}  we note that the lower bounds \eqref{eqn:LowerBoundCap1b} in Prop. \ref{pro:LowerBoundCap} also lower bound the achievable rate of the \ac{lgmc} with {\em stationary multivariate Gaussian input}. This implies that \eqref{eqn:LowerBoundBB1b} constitutes a lower bound on the achievable rate of  \ac{bb}-\ac{plc} channels with cyclostationary Gaussian input. Consequently, when \eqref{eqn:LowerBoundBB1b} coincides with the upper bound in \eqref{eqn:UpperBoundBB1}, then cyclostationary Gaussian inputs are optimal.
	}
\end{remark}

\vspace{-0.4cm}
\section{Application:  Capacity Bounds for Several \ac{bb}-\ac{plc} Channel Models} 
\label{sec:SecBBPLC}
\vspace{-0.1cm}
The capacity bounds derived in Section \ref{sec:SecCap} depend on the marginal distribution of the noise in the \ac{bb}-\ac{plc} channel, $\Weq[i]$, only through its entropy rate.
In this section we derive explicit expressions for the entropy rates of two common non-Gaussian \ac{bb}-\ac{plc} noise models: The Nakagami-$m$ model \cite{Meng:05}, and the \ac{gm}\footnote{The Middleton class A  distribution, which is another important \ac{bb}-\ac{plc} noise model, can be  approximated using a \ac{gm} distribution \cite{Gianaroli:14},  thus the entropy rate of a Middleton class A noise can be approximated using the entropy rate of a \ac{gm}  process.} model \cite{Gianaroli:14}.
%
We first consider the case in which the noise is an i.i.d. process, and thus its entropy rate is equal to the differential entropy of a single sample \cite[Ch. 4.2]{Cover:06}. In such cases, the entropy rate of the noise process can be computed using only the marginal distribution of the noise.
When the noise is correlated, then the derivation of the entropy rate requires the  characterization of the complete statistics of the noise process, which is typically unavailable for the current \ac{bb}-\ac{plc} channel models. Thus, in this work we incorporate  periodically time varying noise autocorrelation functions by applying \ac{lptv} filtering to an i.i.d. noise process, 
and using the entropy rate of the resulting output process in our expressions.
In order to apply this approach, we first  derive a relationship between the entropy rates at the input and at the output of LPTV filters, when the input is an i.i.d. process. We note, however, that for non-Gaussian processes, \ac{lptv} filtering typically does not preserve the marginal distribution of the input process at the output, hence the resulting output process will typically have a mismatched marginal distribution w.r.t. that of the input process. Accordingly, the bounds obtained using the proposed approach should be considered as an indication of the bounds on the capacity of BB-PLC channels with correlated noise. 
In the following  we propose exact expressions and bounds on  the entropy rate $\HWiBB$. These expressions and bounds can be used in Corollaries \ref{cor:UpperBoundBB} and \ref{cor:LowerBoundBB} to obtain bounds on the capacity of several \ac{bb}-\ac{plc} models.

\vspace{-0.2cm}
\subsection{i.i.d. Complex Nakagami-$m$ Noise}
\label{subsec:IID}
\vspace{-0.1cm}
The complex Nakagami-$m$ noise model is a model for the additive noise in baseband \ac{bb}-\ac{plc} channels \cite{Meng:05},  accommodated by our real multivariate model \eqref{eqn:RxModel_1} by representing complex signals  using real multivariate signals.
To facilitate the introduction of this noise model, we recall the definition of the real-valued Nakagami-$m$ distribution:
\begin{definition}[Real-valued Nakagami-$m$ distribution]
	\label{def:Rea; Nakagami-m}
	A real-valued scalar \ac{rv} is said to follow a {\em Nakagami-$m$} distribution with shape parameter $m \ge \frac{1}{2}$ and second-order moment $\Omega >0$  if its \ac{pdf} is given by \cite[Ch. 4.18]{Michalowicz:13}
	  \vspace{-0.1cm}
	\begin{equation}
	\label{eqn:RNekagaemyPDF2}
	\pdf{X}{x}\! = \!\frac{2}{\Gamma\left(m\right)}\left( \frac{m}{\Omega}\right)^mx^{2m-1}e^{-\frac{mx^2}{\Omega}}, \qquad x\ge 0,
	  \vspace{-0.1cm}
	\end{equation}
	where $\Gamma(\cdot)$ denotes the Gamma function. We denote this distribution with 
	$X\sim\Nakagami{m}{\Omega}$.
\end{definition}
The real-valued Nakagami-$m$ distribution is commonly used to model the distribution of the {\em amplitude} of the noise in baseband \ac{bb}-\ac{plc} channels \cite{Meng:05, Mathur:14, Mathur:15}, for which the  marginal distribution of the baseband noise is a complex-valued Nakagami-$m$ PDF \cite{Meng:05}, defined as follows:
\begin{definition}[Complex-valued Nakagami-$m$ distribution \cite{Meng:05}]
\label{def:Complex Nakagami-m}
Let $X\sim\Nakagami{m}{\Omega}$, and let $\Theta$ be an \ac{rv} uniformly distributed over $[{0},{2\pi}]$, mutually independent of $X$. Then,  $\WW=Xe^{j\Theta}$ is a {\em complex Nakagami-$m$} \ac{rv} with zero mean and variance $\Omega$, and is denoted by  $\WW\sim\CNakagami{m}{\Omega}$.
\end{definition}
Letting  $\Psi(\cdot)$ denote the Digamma function \cite[Tbl. 0.1]{Michalowicz:13}, the differential entropy of a complex Nakagami-$m$ \ac{rv} is stated in the following proposition:
\begin{proposition}
\label{thm:NakagamiEntropy}
The differential entropy of $\WW\sim\CNakagami{m}{\Omega}$ is given by:
  \vspace{-0.1cm}
 \begin{equation}
 \label{eqn:NakagamiEntropy}
 h\left( \WW\right) =\frac{1}{2\ln(2)}\Psi\left( m\right)+\log\left(\frac{\pi\Omega}{m}\Gamma\left(m\right)e^{\frac{2m-(2m-1)\Psi\left(m \right) }{2}} \right)
   \vspace{-0.1cm}
 \end{equation}

 \noindent
 [A proof is given in Appendix \ref{app:NakagamiEntropy}]
\end{proposition}
Now, for scalar BB-PLC channels in which the noise is modeled as an i.i.d. complex Nakagami-$m$ process, the entropy rate $\HWiBB$ is given by \eqref{eqn:NakagamiEntropy}, which can be used in \eqref{eqn:UpperBoundBB1}-\eqref{eqn:LowerBoundCBB1} to obtain upper and lower bounds on the capacity.

\vspace{-0.2cm}
\subsection{i.i.d. Gaussian Mixture Noise}
\label{subsec:IIDGMM}
\vspace{-0.1cm}
Next, we consider an additive multivariate real-valued \ac{gm} noise, which is another  common  model for \ac{bb}-\ac{plc} noise, see, e.g.,  \cite{Gianaroli:14}. This model is again obtained by  representing the complex-valued baseband channel as a real-valued channel of extended dimensions. 

Let $\mvn{\ui}{\mean{}}{\var{}}{\tnr}$ denote the \ac{pdf} of an $\tnr\times 1$ real Gaussian random vector with mean vector $\mean{}\in\mathcal{R}^{\tnr}$ and covariance matrix $\var{}\in\mathcal{R}^{\tnr\times \tnr}$, where $\ui$ denotes the realization of the random vector, i.e.,
$\mvn{\ui}{\mean{}}{\var{}}{\tnr} = |2\pi\var{}|^{-1/2}e^{-(\ui - \mean{})^T\var{}^{-1}(\ui - \mean{})}$.
The distribution of a \ac{gm} random vector $\Wi\in\mathcal{R}^{\tnr}$ is determined by the number of Gaussians $\nG$, $\nG\geq 1$, the set of positive mixing parameters $\left\lbrace \prior{n}\right\rbrace_{n=1}^{\nG}$ satisfying $\sum\limits_ {n=1}^{\nG}\prior{n} = 1$, the set of mean vectors $\{\mean{n}\}_{n=1}^{\nG}$, and the set of covariances matrices  $\left\lbrace \var{n}\right\rbrace_{n=1}^{\nG}$. Using these parameters, the \ac{pdf} of $\Wi$ is given by:
 \vspace{-0.1cm}
\begin{equation}
\label{eqn:GMpdf}
\pdf{\Wi}{\wi}=\sum_{n=1}^{\nG}\prior{n}\cdot \mvn{\wi}{\mean{n}}{\var{n}}{\tnr}.
 \vspace{-0.1cm}
\end{equation}
While there is no closed-form analytic expression for the differential entropy of \ac{gm} random vectors \cite{Huber:08}, upper and lower bounds on the differential entropy of \ac{gm} random vectors can be obtained as stated in \cite[Thm. 2-3]{Huber:08}, repeated here for convenience:
\begin{theorem*}
	{\em \cite[Thms. 2-3]{Huber:08}}.
	The differential entropy of a random vector with \ac{pdf} \eqref{eqn:GMpdf} satisfies
	\begin{equation*}
	-\sum_{n=1}^{\nG}\prior{n}\cdot\log\left(\sum_{m=1}^{\nG}\prior{m}\cdot\mvn{\mean{n}}{\mean{m}}{\var{m}\!+\!\var{n}}{\tnr} \right)\leq h(\Wi)\leq\sum_{n=1}^{\nG}\prior{n}\cdot\Big(\frac{1}{2}\log  |2\pi e\var{n}| -\log\left( \prior{n}\right)  \Big).
	\end{equation*}
\end{theorem*}

The bounds in \cite[Thms. 2-3]{Huber:08} are tight when the number of Gaussian components is small\footnote{In the case of $\nG=1$, i.e., a multivariate Gaussian distribution, the upper bound is the differential entropy.} and when the  Gaussians are well separated from each other \cite{Huber:08}, which applies to the \ac{gm} \ac{bb}-\ac{plc} noise model in \cite{Gianaroli:14}. As for i.i.d. noise  $\HWiBB= h\left( \Wi\right)$, \cite[Thms. 2-3]{Huber:08} provide tight bounds on the entropy rate of i.i.d. \ac{gm} noise for small $\nG$ and sufficiently separated Gaussians.

%

\vspace{-0.2cm}
\subsection{Correlated Non-Gaussian Cyclostationary Noise}
\label{subsec:Colored}
\vspace{-0.1cm}
In the previous subsections we studied the differential entropy of two i.i.d. \ac{bb}-\ac{plc} noise models. As in many  \ac{bb}-\ac{plc} systems the noise process is modeled as a temporally correlated \cite{Gianaroli:14, Gotz:04, Tonello:14, Esmailian:03, Galli:11} cyclostationary process \cite{Gotz:04, Corripio:06, Cortes:10}, we propose an approach for extending the  derivation of the differential entropy for i.i.d. noise models studied in Subsections \ref{subsec:IID}--\ref{subsec:IIDGMM} to correlated non-Gaussian cyclostationary noise models.

\label{txt:EntRate}
In order to compute the capacity bounds  in \eqref{eqn:UpperBoundBB1}-\eqref{eqn:LowerBoundCBB1}, it is required to compute $\frac{1}{\Per}\HWiBB$, which is the entropy rate of the multivariate noise process $\Wi\BB\left[\,\ieq\,\right]$.
As the noise $\Weq[i]$ is a temporally and spatially correlated non-Gaussian cyclostationary process, then computing the entropy rate of $\Wi\BB\left[\,\ieq\,\right]$ requires the complete  statistics of the noise process. We note, however, that complete statistical models for the noise in \ac{bb}-\ac{plc} channels are currently not available for most typical \ac{bb}-\ac{plc} scenarios \cite{Gianaroli:14}.  
In the following we apply the widely acceptable practice of generating a correlated noise process via filtering an appropriate i.i.d. process. 
Accordingly, we propose to obtain an explicit expression for the entropy rate by modeling the noise process as the output of an \ac{lptv} filter with an i.i.d. non-Gaussian input.
This model accounts for the non-Gaussianity of the noise, as well as for its cyclostationarity, temporal correlation, and spatial correlation.
We note that the approach has been applied previously in the context of noise generation for narrowband \ac{plc} systems in \cite{Evans:12, IEEE:11}.
%
The noise signal is generated as described below: First, we let $\Ueq[i] \in \mathcal{R}^{\tnr}$ be an i.i.d. random process, and let $\big\{\tilde{\Fi}[i,\tau]\big\}_{\tau=0}^{\Memeq}$ be the \ac{cir} of an $\tnr \times \tnr$ \ac{lptv} filter with period $\Per$ and memory $\Memeq$, where $\tilde{\Fi}[i,0]$ is non-singular $\forall i \in \tilde{\PerSet}$. The noise process is then generated via
\vspace{-0.1cm}
\begin{equation}
\label{eqn:LPTVFilt1}
\Weq\left[i\right] = \sum\limits_{\tau = 0}^{\Memeq}\tilde{\Fi}[i,\tau] \Ueq\left[i - \tau\right].
\vspace{-0.1cm}
\end{equation}
Note that the resulting noise process $\Weq\left[i\right]$ is a strict-sense cyclostationary process with a period of $\Per$ samples and a temporal correlation which spans an interval of $\Memeq$ samples, hence it satisfies the model assumptions in Subsection \ref{subsec:Pre_Model}.

We next consider blocks of $\Per \cdot \tnr$ samples of  $\Weq\left[i\right]$, and restate the \ac{lptv} filtering of \eqref{eqn:LPTVFilt1} as a multivariate \ac{lti} filtering of extended dimensions. To that aim, define the $\Per \cdot \tnr  \times \Per \cdot \tnr$ matrices $\Fi[0]$ and $\Fi[1]$:
\vspace{-0.2cm}
\begin{equation*}
\Fi[0] \!\triangleq\! \left[ {\begin{array}{*{20}{c}}
	{\tilde{\Fi}\!\left[0,0\right]}&\cdots &0&  \cdots &0\\
	\vdots & \ddots &{}&\ddots& \vdots \\
	{\tilde{\Fi}\!\left[\Memeq,\Memeq\right]}& \cdots &{\tilde{\Fi}\!\left[\Memeq,0\right]}& \cdots &0\\
	\vdots & \ddots &{}& \ddots & \vdots\\
	0& \cdots &{\tilde{\Fi}\!\left[\Per\! - \!1,\Memeq\right]}& \cdots &{\tilde{\Fi}\!\left[\Per\! - \!1,0\right]}
	\end{array}} \right], \quad
\Fi[1] \!\triangleq\! \left[ {\begin{array}{*{20}{c}}
	0& \cdots &0&{\tilde{\Fi}\!\left[0,\Memeq\right]}& \cdots &{\tilde{\Fi}\!\left[0,1\right]}\\
	\vdots &{}& \vdots &{}& \ddots & \vdots \\
	0& \cdots &0&0&{}&{\tilde{\Fi}\!\left[\Memeq\! - \!1,\Memeq\right]}\\
	\vdots &{}& \vdots & \vdots &{}& \vdots \\
	0& \cdots &0&0& \cdots &0
	\end{array}} \right],
\end{equation*}
and let $\Ffd(\omega) \!\triangleq \! \sum\limits_{\tau=0}^1\!\Fi[\tau]e^{-j\omega \tau}$.
Also, recall that
$\Wi\BB\left[\,\ieq\,\right] \triangleq \Weq_{\ieq \cdot \Per}^{\left(\ieq \! + \! 1\right)\cdot \Per - 1}$ and let $\Ui\left[\,\ieq\,\right] \triangleq \Ueq_{\ieq \cdot \Per}^{\left(\ieq \! + \! 1\right)\cdot \Per - 1}$.
From \eqref{eqn:LPTVFilt1} we obtain the following relationship between  $\Wi\BB\left[\,\ieq\,\right]$ and $\Ui\left[\,\ieq\,\right]$:
\vspace{-0.1cm}
\begin{equation}
\label{eqn:LPTVFilt2}
\Wi\BB\left[\,\ieq\,\right] = \sum\limits_{\teq = 0}^{1}\Fi[\teq] \Ui\left[\,\ieq - \teq\right].
\vspace{-0.1cm}
\end{equation}
Since $\Ueq[i]$ is an i.i.d. process, it follows that the entropy rate of $\Ui\left[\,\ieq \,\right]$ is given by $\Per \cdot h\big(\Ueq\big)$.
We can now obtain the time-averaged entropy rate of $\Wi\BB\left[\,\ieq\,\right]$ as stated in the following lemma:
\begin{lemma}
	\label{cor:entropyGain}
	The time-average of the entropy rate of $\Wi\BB\left[\,\ieq\,\right]$ is given by
	\vspace{-0.1cm}
	\begin{equation}
	\label{eqn:EntropyGainBB}
	\frac{1}{\Per}\HWiBB=\frac{1}{2\pi\cdot \Per}\int\limits_{\omega=0}^{2\pi}\log \left| \Ffd\left( \omega\right) \right| d\omega +  h\big(\Ueq \big).
	\vspace{-0.1cm}
	\end{equation}

	\noindent
	[A proof is given in Appendix \ref{app:EntropyGain}]	
\end{lemma}

We note that modeling the noise via \eqref{eqn:LPTVFilt1} allows us to evaluate the entropy rate for non-Gaussian, temporally correlated, and cyclostationary  \ac{bb}-\ac{plc} noise models.
It should be noted that, in general, the marginal distribution of $\Weq[i]$ may be different than the marginal distribution of the i.i.d. signal $\Ueq[i]$, e.g., when $\Ueq[i]$ follows a complex Nakagami-$m$ distribution, 
yet,  when $\Ueq[i]$ is a \ac{gm} process, then  the filtered process in \eqref{eqn:LPTVFilt1} is also a \ac{gm} process, but the number of Gaussians and their parameters may change \cite{Schrempf:05}.

\vspace{-0.25cm}
\section{Numerical Examples and Discussion}
\label{sec:Simulations}
\vspace{-0.05cm}
In this section we numerically evaluate the capacity bounds derived in Section \ref{sec:SecCap} for various \ac{bb}-\ac{plc} channels. The simulation study consists of two parts:
First, in Subsection \ref{subsec:SimNonGaussEffect} we illustrate the effect of the non-Gaussianity of the noise on the capacity of the  channel.
Then, in Subsection \ref{subsec:SimBBPLC} we evaluate the capacity bounds for some \ac{bb}-\ac{plc} channel models, considering both scalar as well as \ac{mimo} models, and discuss the tightness of these bounds.

To compute the capacity bounds for the \ac{gm} noise, we first compute upper and lower bounds on the differential entropy $\HWiBB$, denoted $\HWiBB^{\rm (up.)}$ and $\HWiBB^{\rm (low.)}$, respectively, as detailed in Subsection \ref{subsec:IIDGMM}. Then, we compute the upper bound in \eqref{eqn:UpperBoundBB1} by replacing $\frac{1}{\Per}\HWiBB$ with $\frac{1}{\Per}\HWiBB^{\rm (low)}$, and the lower bound in \eqref{eqn:LowerBoundBB1b} (denoted {\em Lower bound 2}) is computed with $\frac{1}{\Per}\HWiBB$ replaced with\footnote{Since for $a,b >0$, the function $f(x) = a\cdot\log\big(b+2^{x/a}\big) - x$ is monotonically non-decreasing w.r.t. $x$, then, computing  \eqref{eqn:LowerBoundBB1b} with  $\frac{1}{\Per}\HWiBB^{\rm (low.)}$ instead of  $\frac{1}{\Per}\HWiBB$ results in a lower bound on the capacity.}  $\frac{1}{\Per}\HWiBB^{\rm (up.)}$. Lastly, we note that  the lower bound in \eqref{eqn:LowerBoundBB1a} (denoted {\em Lower bound 1}) does not depend on the entropy rate.
As \ac{bb}-\ac{plc} channels exhibit a broad range of signal attenuation and noise power values, depending on the topology of the power line network  and on the appliances connected to the network \cite{Gotz:04, Tonello:14, Meng:05,Corripio:06}, we consider a wide range of \ac{snr} values.

\vspace{-0.2cm}
\subsection{Evaluating the Effect of the non-Gaussianity of the Noise}
\label{subsec:SimNonGaussEffect}
\vspace{-0.1cm}
As noted in Section \ref{sec:Intro}, previous works on the fundamental rate limits of \ac{bb}-\ac{plc} channels, e.g., \cite{Tonello:14, Lampe:13, Bo:07}, assumed that the additive noise is Gaussian, which facilitated obtaining an explicit expression for the capacity.
Nonetheless, \ac{bb}-\ac{plc} noise is typically modeled as a {\em non-Gaussian process}, and two common models for its marginal \ac{pdf} are the Nakagami-$m$ distribution \cite{Meng:05} and the \ac{gm} distribution \cite{Gianaroli:14}.
In the following we illustrate the effect of the non-Gaussianity of the additive \ac{bb}-\ac{plc} noise on the capacity of the channel, and numerically evaluate the mismatch induced by assuming that the noise is Gaussian (e.g., as done in some previous works, including \cite{Tonello:14, Lampe:13}) compared to the actual capacity.
To that aim, we consider a memoryless ($\Memeq=1$), time-invariant ($\Per = 1$), scalar baseband channel, in which the additive noise $\Weqs[i]$ is an i.i.d. process. We consider two marginal distributions of the noise:
The first noise process follows a complex \ac{gm} distribution. In this case, in order to generate $\Weqs[i]$, we let $ \tilde{Z}[i]$ be an i.i.d. complex process such that $\Big[{\rm Re}\big\{\tilde{Z}[i]\big\}, {\rm Im}\big\{\tilde{Z}[i]\big\}\Big]^T$ is a $2\times1$ \ac{gm} random vector with parameters  $\nG=3$,  $\{\prior{n}\} _{n=1}^{3} = \{0.7, 0.2, 0.1\}$, $\{\mean{n}\}_{n=1}^{3}= \{[5,4]^T, [-8,-16]^T, [-19,4]^T \}$, and $\{\var{n}\}_{n=1}^{3} = \{5,2,1\}\cdot \myMat{I}_2$, following \cite[Fig. 3a]{Gianaroli:14}. Then, we set $\alpha = \E\{|\tilde{Z}[i]|^2\}^{-1/2}$, and obtain the noise as $\Weqs[i] = \alpha \cdot \tilde{Z}[i]$;
We also consider noise with a complex Nakagami-$m$ distribution with parameters $m=0.8$ and $\Omega=1$, as in \cite{Meng:05}.
Note that both noise models have a zero mean and a unit variance.
This scenario accounts only for the non-Gaussianity of the noise in the channel model, and neglects the effects of the channel memory and of the non-stationarity of the noise.

\begin{figure}
   	\centering
   	\begin{minipage}{0.45\textwidth}
   		\centering
   		\scalebox{0.48}{\includegraphics{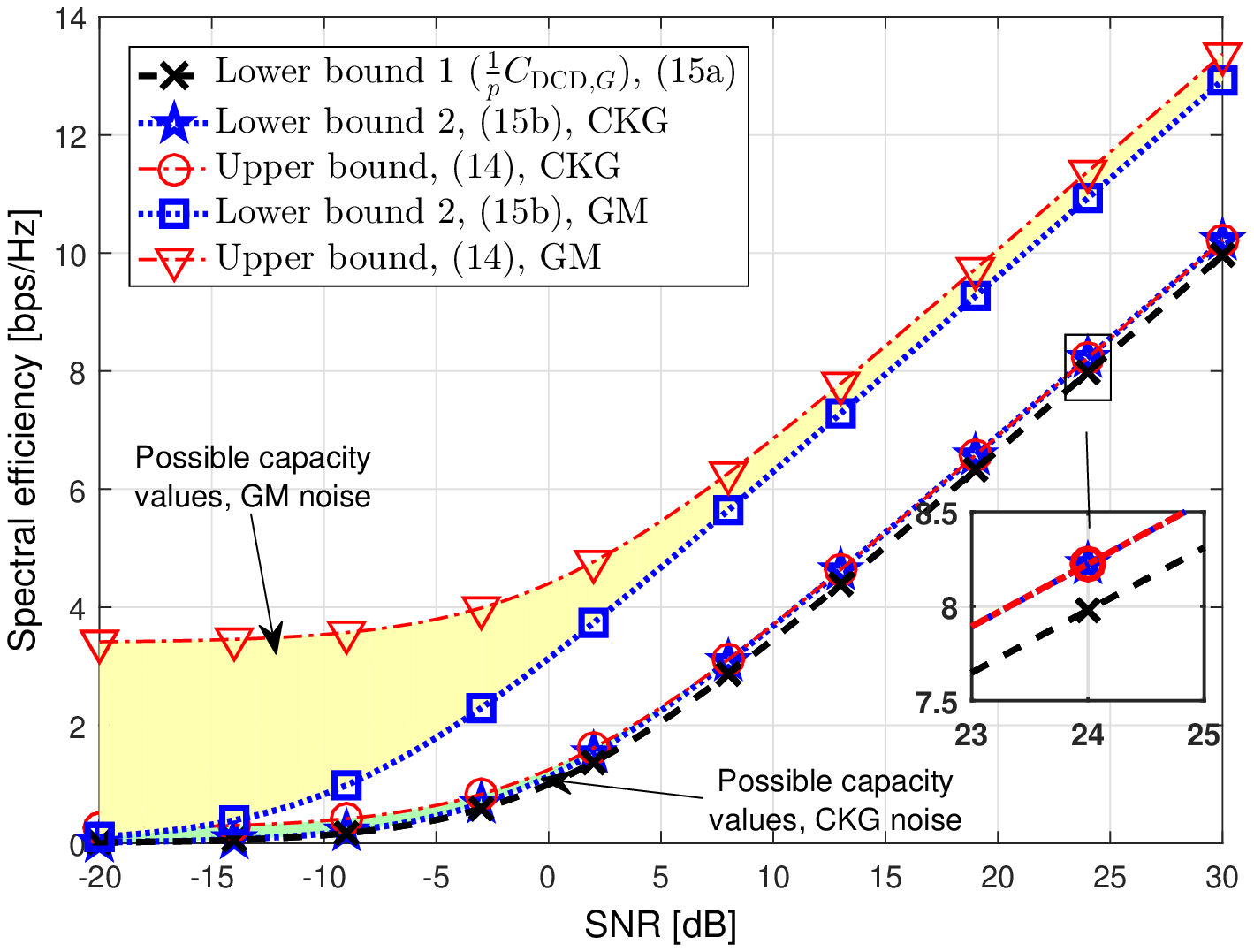}}
   		\vspace{-0.8cm}
   		\caption{Capacity bounds for the i.i.d. noise channel, with complex Nakagami-$m$ (CKG) and \ac{gm} noise models.
   		}
   		\label{fig:iid_noise}		
   	\end{minipage}
   	$\quad$
   	\begin{minipage}{0.45\textwidth}
   		\centering
   		\scalebox{0.48}{\includegraphics{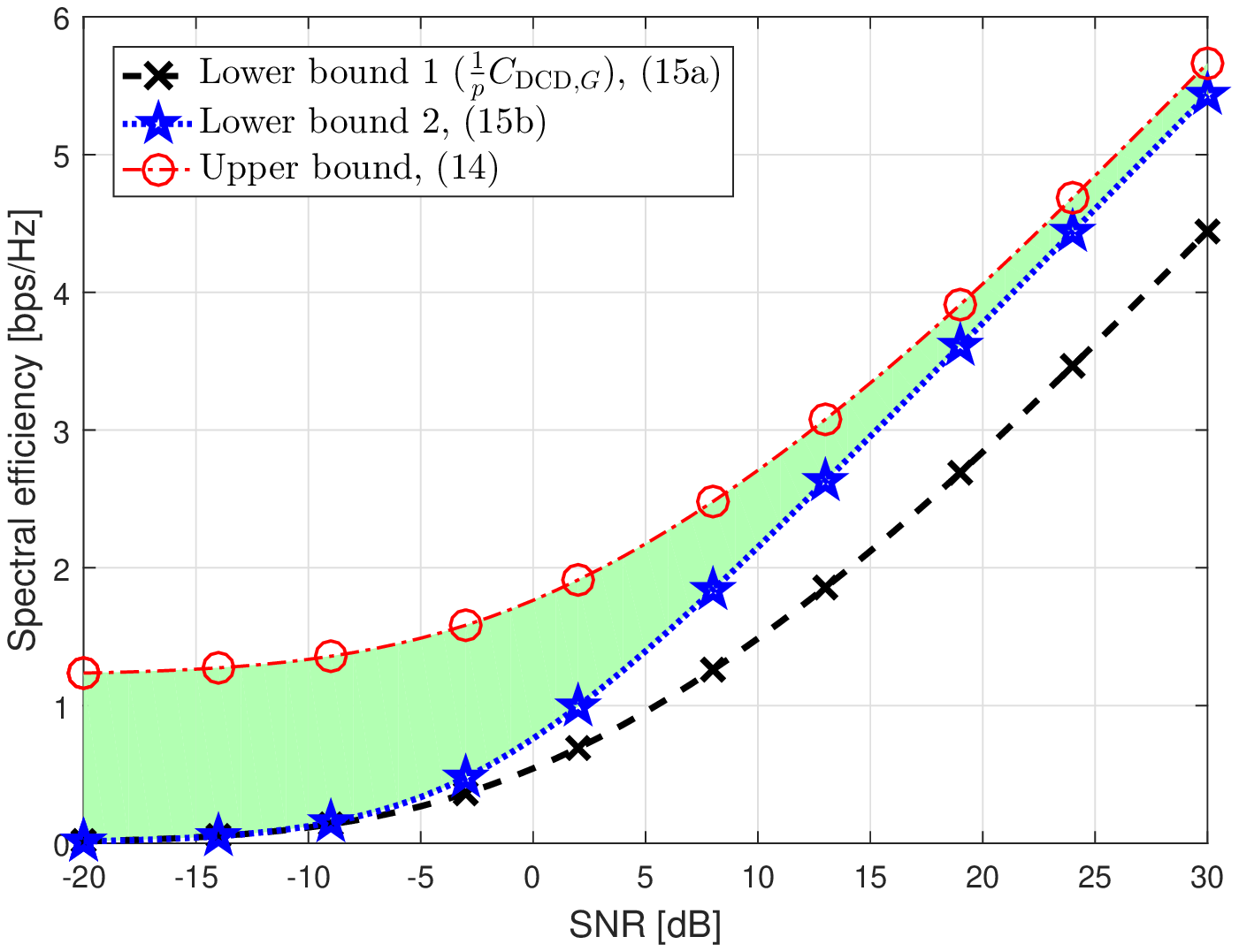}}
   		\vspace{-0.8cm}
   		\caption{Capacity bounds for the scalar \ac{bb}-\ac{plc} channel, with the {\bf GM1} noise model.
   		}
   		\label{fig:PLC_Scalar_noise}
   	\end{minipage}
   	\vspace{-0.9cm}
   \end{figure}

Fig. \ref{fig:iid_noise} depicts the capacity bounds for this scenario vs. \ac{snr}, defined here as ${\rm SNR} = \frac{\PCsteq} { \E\{|\Weqs[i]|^2\}}$.
Note that the lower bound in \eqref{eqn:LowerBoundBB1a} ({\em Lower bound 1} in Fig. \ref{fig:iid_noise}), which represents the capacity of the channel assuming that the noise is Gaussian, does not depend on the actual distribution of the noise, and is therefore the same for both simulated noise distributions.
Observing Fig. \ref{fig:iid_noise}, we note that for \ac{gm} noise,  there is a substantial gap between the actual capacity of the channel and the capacity computed assuming that the noise is Gaussian, especially in high \ac{snr}. For example, at \ac{snr} of $12$ dB, capacity is not less than $7$ bps/Hz, while assuming Gaussian noise, the \ac{snr} has to be increased by at least $9$ dB in order to obtain the same capacity of  $7$ bps/Hz. A less notable gap is observed for  Nakagami-$m$ noise, where an \ac{snr} gap of $0.7$ dB is observed for capacity of $7$ bps/Hz.
Moreover, we note that for the  Nakagami-$m$ noise, the lower bound in \eqref{eqn:LowerBoundBB1b} numerically coincides with the upper bound for \acp{snr} greater than $10$ dB. As discussed in Comment \ref{rem:CorBB_1}, this implies that  Gaussian inputs are {\em optimal} at high \ac{snr}  for the  Nakagami-$m$ noise channel. For the \ac{gm} noise model, we observe a gap of $0.5$ bps/Hz between the lower bound  \eqref{eqn:LowerBoundBB1b} and the upper bound \eqref{eqn:UpperBoundBB1}, for \acp{snr} above $10$ dB.
Consequently, as \eqref{eqn:LowerBoundBB1b} lower bounds the achievable rate with Gaussian inputs, we conclude that for the \ac{gm} noise model, the achievable rate of Gaussian inputs is at most $0.5$ bps/Hz less than capacity at high \ac{snr}.
%

\vspace{-0.3cm}
\subsection{Capacity of \ac{bb}-\ac{plc} Channels with Correlated Non-Gaussian Noise}
\label{subsec:SimBBPLC}
\vspace{-0.1cm}
We now use the results in Corollaries \ref{cor:UpperBoundBB} and \ref{cor:LowerBoundBB} to characterize bounds on the capacity of practical \ac{bb}-\ac{plc} channel models. 
The channel models considered here are taken from the recent literature on \ac{bb}-\ac{plc} channel modeling, and are selected to represent actual \ac{bb}-\ac{plc} channels.
We first study the capacity of the scalar passband \ac{bb}-\ac{plc} scenario: 
The \ac{lptv} \ac{cir} is generated with period a $\PerCh = 240$ and memory length\footnote{Note that the \ac{rms} delay spread in \ac{bb}-\ac{plc} channels is typically on the order of several microseconds, i.e., around $0.1 \%$ of the channel period \cite[Tbl. 1]{Berger:15}. Thus, following the typical relationship between \ac{rms} delay spread and memory length, see, e.g., \cite[Ch. 3.3.1]{Goldsmith:05}, the memory length is on the order of $1 \%$ of the channel period.} $\Memeq = 4$ using the channel generator proposed in \cite{Corripio:11}, where the parameters used by the channel generator were set to the default values.
The additive noise is a non-Gaussian temporally correlated cyclostationary process, generated using the approach described in Subsection \ref{subsec:Colored}:
First, an i.i.d. scalar process $\tilde{U}[i]$ is generated, where we consider three  \acp{pdf} for $\tilde{U}[i]$:
\begin{itemize}
	\item {\bf GM1} - a \ac{gm} \ac{pdf} based on \cite[Fig. 3a]{Gianaroli:14} 	with parameters  $\nG=3$,  $\{\prior{n}\} _{n=1}^{3} = \{0.7, 0.2, 0.1\}$, $\{m_n\}_{n=1}^{3}= \{5, -8, -19 \}$, and $\{c_n\}_{n=1}^{3}= \{5, 2, 1 \}$; 
	\item {\bf GM2} - a \ac{gm} \ac{pdf} based on \cite[Fig. 2]{Lin:13}  	with parameters  $\nG=3$,  $\{\prior{n}\} _{n=1}^{3} = \{0.9, 0.07, 0.03\}$, $\{m_n\}_{n=1}^{3}= \{0, 0, 0 \}$, and $\{c_n\}_{n=1}^{3}= \{1, 100, 1000 \}$; 
	\item {\bf MCA} - a \ac{gm} \ac{pdf} approximating a Middleton Class A \ac{pdf} as in \cite[Ch. 2.7.2]{Lampe:16} 	with parameters based on \cite[Fig. 3]{Lin:13}, i.e., letting $A = 0.1$ and $\Omega = 0.01$, and setting $\nG=10$,  $\prior{n} = e^{-A} \frac{A^n}{n!}$, $m_n = 0$, and $c_n = \frac{n/A + \Omega}{1 + \Omega}$, $n \in \{0,1,\ldots,\nG-1\}$.
\end{itemize} 
The process  $\tilde{U}[i]$ is normalized to have a unit variance, and is then filtered via a spectral shaping \ac{lptv} filter  to obtain the scalar \ac{bb}-\ac{plc} noise $\Weqs[i]$. Two spectral shaping \ac{lptv} filters with period $\PerN = 120$ and memory length $\Memeq = 4$ are used: The first is a filter designed to generate the periodically time-varying \ac{bb}-\ac{plc} {\em `medium disturbed'} correlation profile. This filter is applied to the {\bf GM1} and {\bf GM2} noise signals. The second spectral shaping filter is  designed to generate the periodically time-varying \ac{bb}-\ac{plc} {\em `heavily disturbed'} correlation profile, and is applied to the {\bf MCA} noise model. Both correlation profiles were obtained from actual \ac{bb}-\ac{plc} noise measurements via the procedure detailed in \cite{Cortes:10}\footnote{The {\em `medium disturbed'} and {\em `heavily disturbed'} correlation profiles obtained following \cite{Cortes:10} are available on \url{http://www.plc.uma.es/channels.htm}.}. Note that for the values selected for $\PerCh$ and $\PerN$, then $\Per$, which is the least common multiple of $\PerCh$ and $\PerN$ not smaller than $\Memeq$, equals $\Per = 240$.

\begin{figure}
   	\centering
   	\begin{minipage}{0.45\textwidth}
   		\centering
   		\scalebox{0.48}{\includegraphics{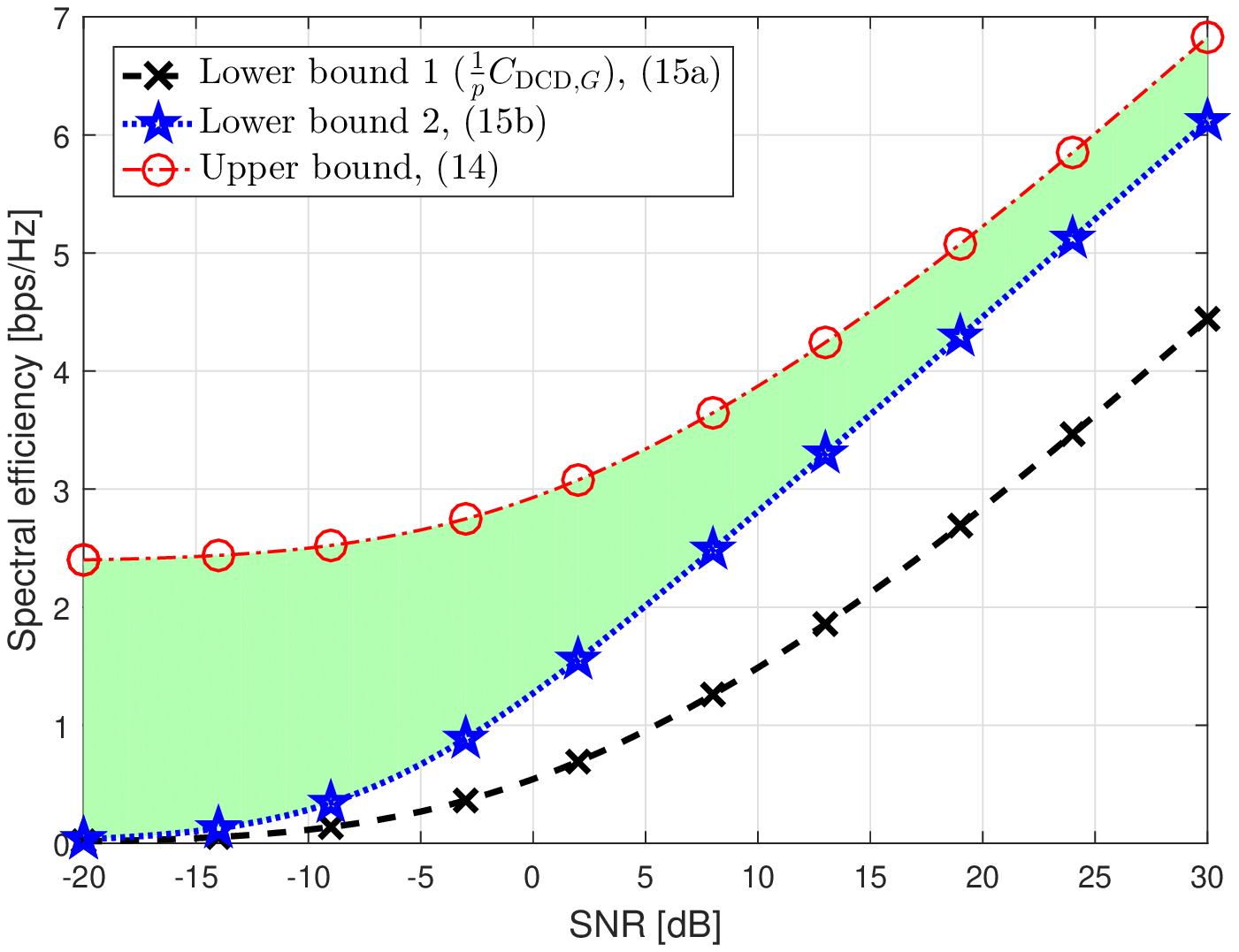}}
   		\vspace{-0.8cm}
   		\caption{Capacity bounds for the scalar \ac{bb}-\ac{plc} channel, with the {\bf GM2} noise model.
   		}
   		\label{fig:PLC_Scalar_GMnoise}		
   	\end{minipage}
   	$\quad$
   	\begin{minipage}{0.45\textwidth}
   		\centering
   		\scalebox{0.48}{\includegraphics{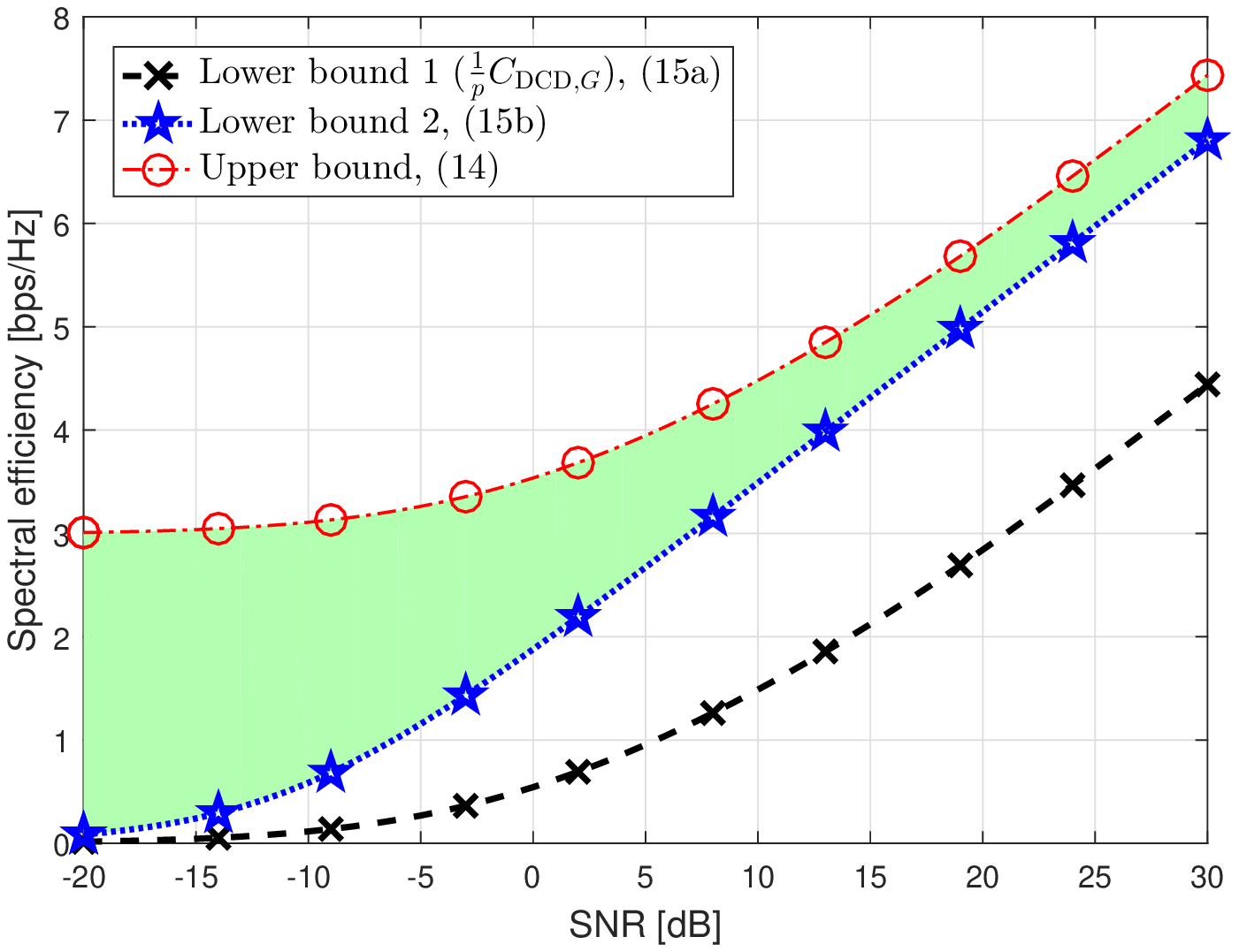}}
   		\vspace{-0.8cm}
   		\caption{Capacity bounds for the scalar \ac{bb}-\ac{plc} channel, with the {\bf MCA} noise model.
   		}
   		\label{fig:PLC_Scalar_MCAnoise}
   	\end{minipage}
   	\vspace{-0.6cm}
   \end{figure}

The capacity bounds for the scalar \ac{bb}-\ac{plc} channel vs. \ac{snr}, defined here as ${\rm SNR} = \frac{\PCsteq} { \frac{1}{\Per}\sum\limits_{i=1}^{\Per}\E\{|\Weqs[i]|^2\}}$, are depicted in Figs. \ref{fig:PLC_Scalar_noise}-\ref{fig:PLC_Scalar_MCAnoise}, for the {\bf GM1} noise, {\bf GM2} noise, and {\bf MCA} noise, respectively. Observing Figs. \ref{fig:PLC_Scalar_noise}-\ref{fig:PLC_Scalar_MCAnoise}, we note that the lower bound in  \eqref{eqn:LowerBoundBB1b} ({\em Lower bound 2} in  Figs. \ref{fig:PLC_Scalar_noise}-\ref{fig:PLC_Scalar_MCAnoise}) is much tighter than the lower bound in \eqref{eqn:LowerBoundBB1a} ({\em Lower bound 1} in  Figs. \ref{fig:PLC_Scalar_noise}-\ref{fig:PLC_Scalar_MCAnoise}) for all the noise models considered. Consequently, assuming that the {\em noise is Gaussian} results in a capacity expression which is {\em strictly smaller} than the actual capacity, and for most \ac{snr} values, this expression is considerably less than the actual capacity.
It thus follows that using the Gaussian noise assumption leads to schemes whose achievable rates are far from achieving the maximal bit rate that can be supported by the \ac{bb}-\ac{plc} channel.
Additionally, we note that for \acp{snr} higher than $10$ dB, the lower bound  \eqref{eqn:LowerBoundBB1b} is lower than the upper bound \eqref{eqn:UpperBoundBB1} by only $0.25$ bps/Hz, $0.8$ bps/Hz, and $0.6$ bps/Hz, for the  the {\bf GM1} noise, the {\bf GM2} noise, and the {\bf MCA} noise, respectively. We conclude that, for the tested scenarios at high \ac{snr}s, the bounds in \eqref{eqn:LowerBoundBB1b} and  \eqref{eqn:UpperBoundBB1} are relatively tight, hence Corollaries \ref{cor:UpperBoundBB} and \ref{cor:LowerBoundBB} provide a reliable characterization of the capacity. We also conclude that at high \acp{snr} the achievable rate obtained with cyclostationary Gaussian inputs is within a small gap from capacity.

Next, we consider a passband $2 \times 2$ \ac{mimo} \ac{bb}-\ac{plc} scenario.
The multivariate \ac{lptv} \ac{cir} $\Heq[i, \tau]$ was generated using the method proposed in \cite{Vernosi:11} for generating \ac{mimo} \ac{bb}-\ac{plc} channels based on the  characteristics of the scalar channel. Specifically, we first generate four real \ac{lptv} \acp{cir} with period $\PerCh = 240$ and memory length $\Memeq = 4$ using the channel generator proposed in \cite{Corripio:11}. We denote the generated channels as $\{\tilde{g}_k[i,\tau]\}_{k=1}^4$. Then, setting $\rho = 0.9$ \cite[Sec. V-B]{Vernosi:11}, the  multivariate \ac{lptv} \ac{cir} is obtained via
\begin{equation*}
\Heq[i, \tau] = {\left[ {\begin{array}{*{20}{c}}
		1\quad&\rho \\
		\rho \quad&1
		\end{array}} \right]^{1/2}}\left[ {\begin{array}{*{20}{c}}
	{{{\tilde g}_1}\left[ {i,\tau } \right]}\quad&{{{\tilde g}_2}\left[ {i,\tau } \right]}\\
	{{{\tilde g}_3}\left[ {i,\tau } \right]}\quad&{{{\tilde g}_4}\left[ {i,\tau } \right]}
	\end{array}} \right]{\left[ {\begin{array}{*{20}{c}}
		1\quad&\rho \\
		\rho \quad&1
		\end{array}} \right]^{1/2}}.
\end{equation*}
The additive multivariate noise  $\tilde{\bf W}[i]$ is generated using the model detailed in Subsection \ref{subsec:Colored}:
First,  a real i.i.d. $2\times 1$  process $\tilde{\bf U}[i]$ is generated, normalized to having a unit variance. We used two different \acp{pdf} for $\tilde{\bf U}[i]$:
\begin{itemize}
	\item {\bf MIMO GM} - a \ac{gm} \ac{pdf}  based on \cite[Fig. 3a]{Gianaroli:14}  with parameters  $\nG=3$,  $\{\prior{n}\} _{n=1}^{3} = \{0.7, 0.2, 0.1\}$, $\{\mean{n}\}_{n=1}^{3}= \{[5,4]^T, [-8,-16]^T, [-19,4]^T \}$, and $\{\var{n}\}_{n=1}^{3} = \{5,2,1\}\cdot \myMat{I}_2$.
	\item {\bf MIMO MCA} - a \ac{gm} \ac{pdf} approximating a Middleton Class A \ac{pdf} as in \cite[Ch. 2.7.2]{Lampe:16} 	with parameters based on \cite[Fig. 3]{Lin:13}, i.e., letting $A = 0.1$ and $\Omega = 0.01$, such that $\nG=10$,  $\prior{n} = e^{-A} \frac{A^n}{n!}$, $\mean{n} = [0,0]^T$, and $\var{n} = \frac{n/A + \Omega}{1 + \Omega}\cdot \myMat{I}_2$, $n \in \{0,1,\ldots,\nG-1\}$.
\end{itemize} 
Next, we generate a spectral shaping multivariate \ac{lptv} filter, $\tilde{\myMat{F}}[i, \tau]$, with period $\PerN = 120$ (i.e., $\Per = 240$) and memory length $\Memeq = 4$,   based on the construction of a spectral correlation profile  for \ac{mimo} \ac{bb}-\ac{plc} channels detailed in \cite{Rende:11}: Let $\rho_{\bf W}(\omega)$ be a $2\pi$-periodic function representing the spectral variations in the spatial correlation. Following \cite[Fig. 5]{Rende:11}, we set $\rho_{\bf W}(\omega) = 0.7 - \frac{|\omega|}{2\pi}$ for $|\omega|< \pi$. Let $s[i, \omega]$  be the instantaneous \acsp{psd}, corresponding to the {\em `heavily disturbed'} profile\footnote{The instantaneous \acsp{psd} are taken from \url{http://www.plc.uma.es/channels.htm}, which is based on \cite{Cortes:10}.}. Lastly, set
\begin{equation*}
\tilde{\myMat{F}}'[i, \omega] = {\left[ {\begin{array}{*{20}{c}}
		1&{{\rho _{\bf{w}}}\left( \omega  \right)}\\
		{{\rho _{\bf{w}}}\left( \omega  \right)}&1
		\end{array}} \right]^{1/2}}\left[ {\begin{array}{*{20}{c}}
	{{s}\left[ {i,\omega } \right]}&0\\
	0&{{s}\left[ {i,\omega } \right]}
	\end{array}} \right]^{1/2}.
\end{equation*}
The \ac{cir} of the multivariate filter  $\tilde{\myMat{F}}[i, \tau]$ is obtained via the inverse Fourier transform
$\tilde{\myMat{F}}[i, \tau] = \frac{1}{2\pi}\int\limits_{\omega=-\pi}^{\pi}  \tilde{\myMat{F}}'[i, \omega] e^{j\omega\tau} d\omega$.
Finally, the additive noise signal $\Weq[i] \in \mathcal{R}^2$ is obtained as the output of $\tilde{\myMat{F}}[i, \tau]$ as in \eqref{eqn:LPTVFilt1}.

\begin{figure}
	\centering
	\begin{minipage}{0.45\textwidth}
		\centering
		\scalebox{0.48}{\includegraphics{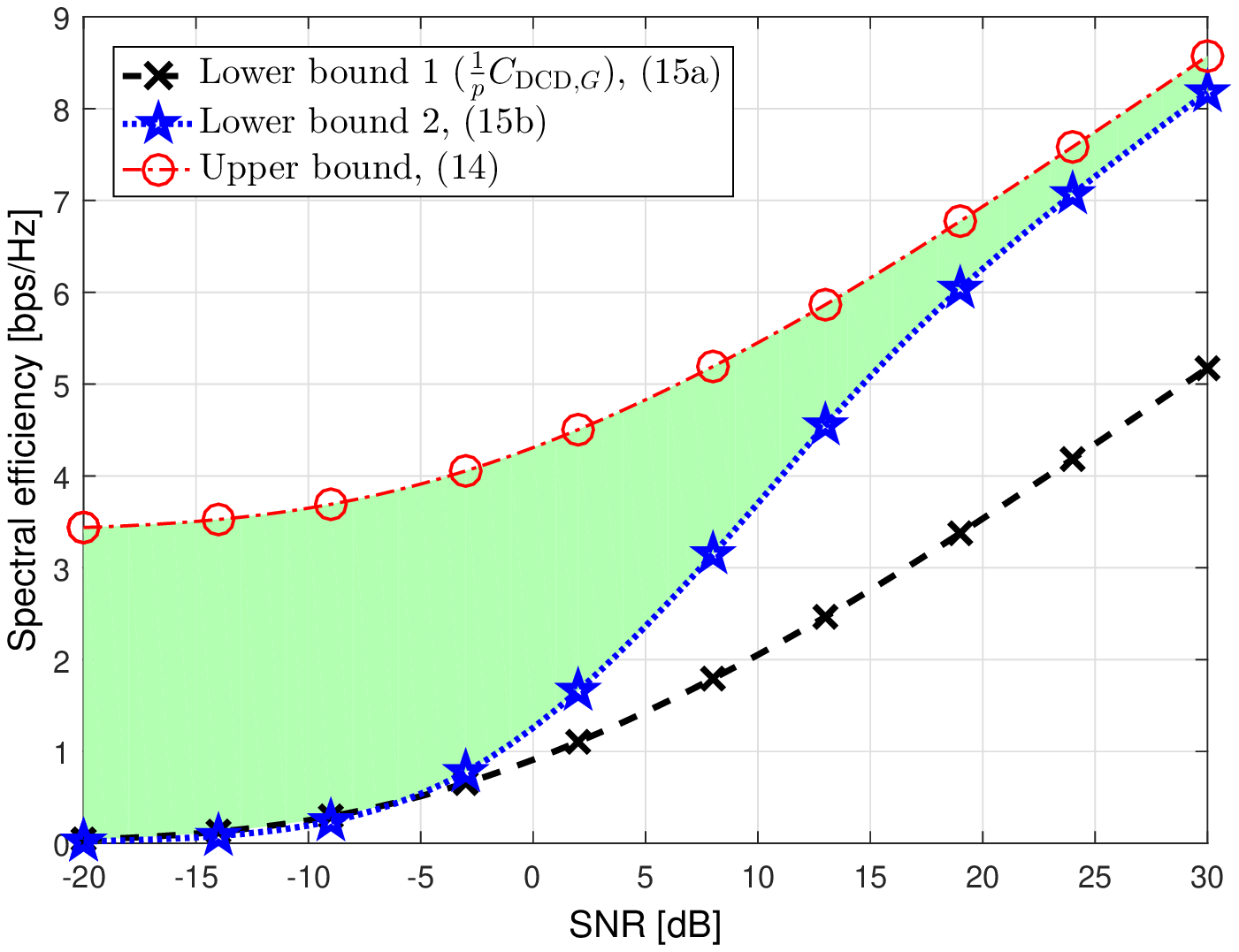}}
		\vspace{-0.8cm}
		\caption{Capacity bounds for the \ac{mimo} \ac{bb}-\ac{plc} channel, with the {\bf MIMO GM} noise.
		}
		\label{fig:PLC_MIMO_noise}		
	\end{minipage}
	$\quad$
	\begin{minipage}{0.45\textwidth}
		\centering
		\scalebox{0.48}{\includegraphics{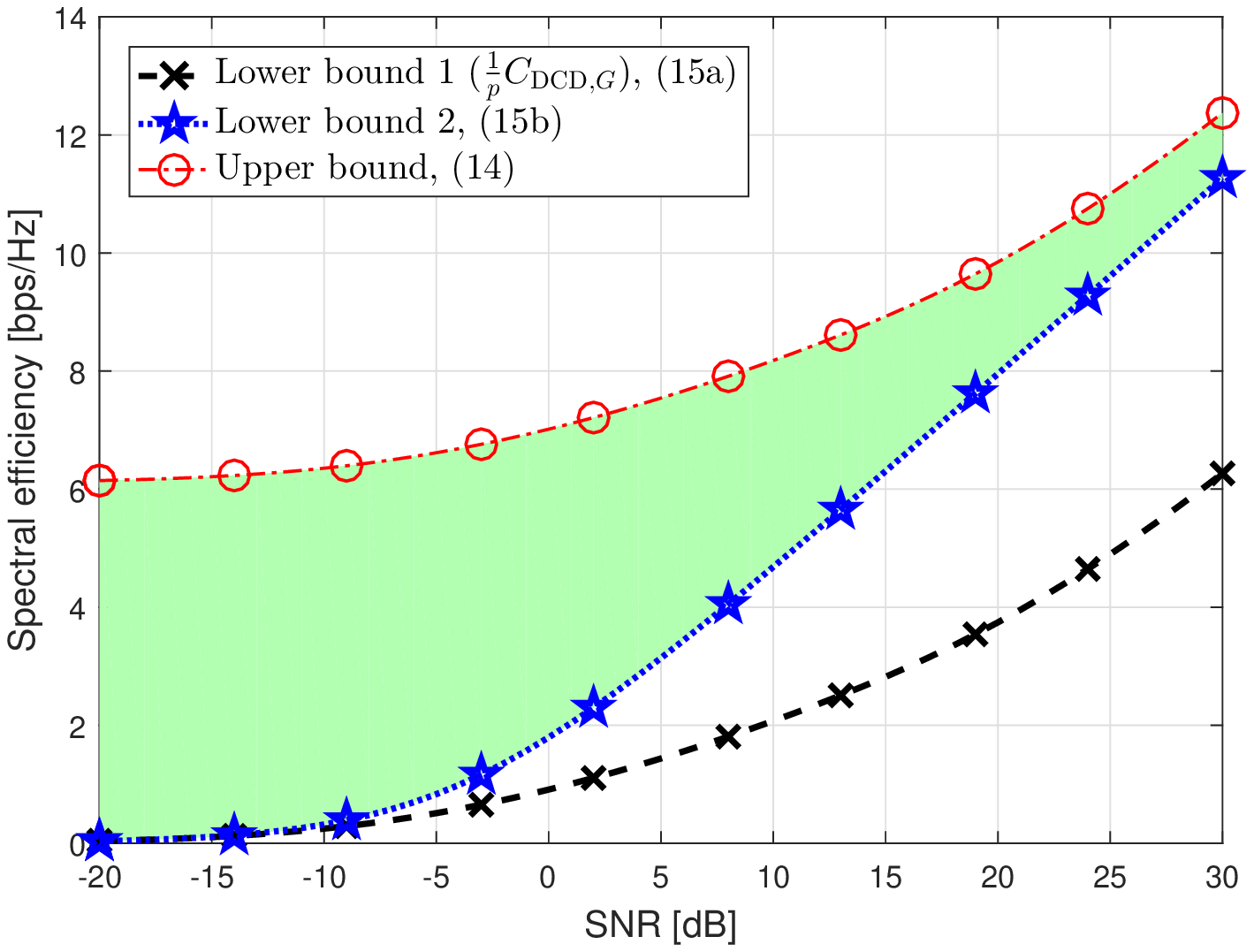}}
		\vspace{-0.8cm}
		\caption{ Capacity bounds for the \ac{mimo} \ac{bb}-\ac{plc} channel, with the {\bf MIMO MCA} noise.
		}
		\label{fig:PLC_MIMO_MCAnoise}
	\end{minipage}
	\vspace{-0.9cm}
\end{figure}

The capacity bounds for the \ac{mimo} \ac{bb}-\ac{plc} channel vs. \ac{snr}, defined here as ${\rm SNR} = \frac{\PCsteq} { \frac{1}{\Per}\sum\limits_{i=1}^{\Per}\E\{\|\Weq[i]\|^2\}}$, are depicted in Figs. \ref{fig:PLC_MIMO_noise}-\ref{fig:PLC_MIMO_MCAnoise} for the {\bf MIMO GM} and for the {\bf MIMO MCA} noise models, respectively. Similarly to the capacity of the scalar \ac{bb}-\ac{plc} channel, the lower bound in \eqref{eqn:LowerBoundBB1b} is tighter than the lower bound in \eqref{eqn:LowerBoundBB1a} for almost the entire \ac{snr} range. We also note that the gap between the tighter lower bound and the upper bound in Figs. \ref{fig:PLC_MIMO_noise}-\ref{fig:PLC_MIMO_MCAnoise} is larger than in the scalar case in Figs. \ref{fig:PLC_Scalar_noise}-\ref{fig:PLC_Scalar_MCAnoise}, varying from $3.05$ bps/Hz at \ac{snr} of $0$ dB to $0.45$ bps/Hz at high \acp{snr} for the {\bf MIMO GM} noise model, while for the {\bf MIMO MCA} noise model the corresponding gap varies from  $4.5$ bps/Hz at \ac{snr} of $0$ dB to $1.1$ bps/Hz at high \acp{snr}.
\label{txt:MIMOComp}
Comparing the capacity of \ac{mimo} \ac{bb}-\ac{plc} channels in Figs. \ref{fig:PLC_MIMO_noise}-\ref{fig:PLC_MIMO_MCAnoise} with their scalar counterparts in Figs. \ref{fig:PLC_Scalar_noise}-\ref{fig:PLC_Scalar_MCAnoise}, respectively, indicates that the potential rate gains of using \ac{mimo} techniques for \ac{bb}-\ac{plc} can range between $40\%-90\%$. Recall that the optimal rate gain of  a $2\times2$ configuration over the scalar channel for spatially independent noise  is $100\%$ \cite[Ch 9]{Tse:05}. Hence, by using two transmit ports and two receive ports, one can achieve gains which are close to the maximal gain. For example, at an \ac{snr} of $20$ dB, we observe in Fig. \ref{fig:PLC_MIMO_noise} that the capacity of the {\bf MIMO GM} noise channel is between $6.2-6.9$ bps/Hz, while for the scalar case, we observe in Fig. \ref{fig:PLC_Scalar_noise} that the capacity is between $3.8-4$ bps/Hz. Thus, the \ac{mimo} configuration can achieve a rate gain of $55\%-81\%$ over the scalar channel. For the {\bf MIMO MCA} noise the corresponding rate gain is $38\%-88\%$. This indicates that \ac{mimo} \ac{bb}-\ac{plc} configurations can achieve significant rate gains over scalar \ac{bb}-\ac{plc} channel at manageable computational complexity \cite[Ch. 7]{Tse:05}, \cite[Ch. 10]{Goldsmith:05}.
Finally, we note that for the considered channel models, it follows from our capacity analysis that a \ac{bb}-\ac{plc} system with a configuration similar to the ITU-T G.9963 standard \cite{ITU:11}, namely, a system which utilizes two transmit ports and two receive ports, over a frequency band of $100$ MHz,   can achieve data rates approaching and even surpassing one Gbps at high \acp{snr}.

\label{txt:Coding}
Our results lead to several insights on practical channel coding for \ac{bb}-\ac{plc} channels: First, observe that at high \acp{snr} for both scalar and \ac{mimo} \ac{bb}-\ac{plc} channels, there is a rather small gap between the achievable rate of cyclostationary Gaussian inputs and capacity. This indicates that at high \ac{snr}, cyclostationary Gaussian codes can closely approach the optimal performance. For lower \ac{snr} values, guidelines to a possible code construction can be obtained from the equivalence between \ac{bb}-\ac{plc} channels and \acp{lgmc}, which belong to the class of time-invariant \ac{mimo} channels, as noted in Subsection \ref{subsec:BBPLC}. Consequently, any code for time-invariant \ac{mimo} channels, can be used in  \ac{bb}-\ac{plc} channels, by applying the inverse \ac{dcd} to the transmitted codeword and the \ac{dcd} to the channel output, achieving the same average probability of error of the code.


\vspace{-0.2cm}
\section{Conclusions}
\label{sec:Conclusions}
\vspace{-0.1cm}
In this paper we characterized upper and lower bounds on the capacity of \ac{mimo} \ac{bb}-\ac{plc} channels, accounting for the unique characteristics of these channels and the non-Gaussianity of the additive noise. We derived capacity bounds which depend on the noise distribution only through its entropy rate and autocorrelation function, and obtained explicit expressions for the entropy rates of several \ac{bb}-\ac{plc} noise models. Our numerical evaluations demonstrate the tightness of the proposed bounds, and illustrate the significant loss resulting from assuming that the noise is Gaussian in the computation of the capacity, which may lead to the design of inherently suboptimal schemes.


\vspace{-0.2cm}
\begin{appendix}

\numberwithin{proposition}{subsection}
\numberwithin{lemma}{subsection}
\numberwithin{corollary}{subsection}
\numberwithin{remark}{subsection}
\numberwithin{definition}{subsection}
\numberwithin{equation}{subsection}

\vspace{-0.2cm}
\subsection{Proof of Proposition \ref{pro:UpperBoundCap}}
\label{app:Proof3}
\vspace{-0.1cm}
In order to prove \eqref{eqn:UpperBoundCap1}, let $\Wi_G[i]$ be a zero-mean Gaussian process with an autocorrelation function $\CWi[\tau]$, defined after \eqref{eqn:EqRelation}, s.t. $\Wi_G[i]$ is mutually independent of the channel input. Note that the mutual information in \eqref{eqn:AsymCapEq} can be written as
\vspace{-0.2cm}
\begin{align}
\frac{1}{n}I\left(\Xin^{n-1} ; \Yi^{n-1} | \Xin_{-\Mem}^{-1} = {\bf 0}_{\nt \cdot \Mem}\right)
&\! = \! \frac{1}{n}h\left( \Yi^{n-1}| \Xin_{-\Mem}^{-1} = {\bf 0}_{\nt \cdot \Mem} \right) - \frac{1}{n}h\left(\Wi^{n-1} \right) \notag \\
&\! = \! \frac{1}{n}\Big( h\left( \Yi^{n-1}| \Xin_{-\Mem}^{-1} = {\bf 0}_{\nt \cdot \Mem} \right) - h\left(\Wi_G^{n-1} \right)\Big) \notag \\
&\qquad   +  \frac{1}{n}h\left(\Wi_G^{n-1} \right)- \frac{1}{n}h\left(\Wi^{n-1} \right).
\label{eqn:AppBeq1}
\vspace{-0.2cm}
\end{align}
Since, for a given correlation function, Gaussian distribution maximizes the differential entropy \cite[Thm. 8.6.5]{Cover:06},   $h\left( \Yi^{n-1}\right)$ is maximized for a Gaussian distribution of $\Yi^{n-1}$ with the same first and second-order moments as the original vector $\Yi^{n-1}$. By letting $\big\{\Yi_G[i]\big\}_{i=0}^{n-1}$ be a Gaussian process with the same first and second-order statistical moments as $\big\{\Yi[i]\big\}_{i=0}^{n-1}$, we have that
\vspace{-0.2cm}
\begin{align}
&\mathop{\lim}\limits_{n \rightarrow \infty} \frac{1}{n}\mathop{\sup}\limits_{\Pdf{\Xin^{n-1}}: \frac{1}{n}\sum\limits_{i\! = \!0}^{n-1}\E \left\{\left\| \Xin\left[ i \right] \right\|^2\right\} \le \PCst  }  h\left( \Yi^{n-1}| \Xin_{-\Mem}^{-1} = {\bf 0}_{\nt \cdot \Mem}\right) - h\left(\Wi_G^{n-1} \right)  \notag \\
&\qquad \stackrel{(a)}{\le} \mathop{\lim}\limits_{n \rightarrow \infty} \frac{1}{n}\mathop{\sup}
\limits_{\Cov\left( \Xin^{n-1}\right)  : {\rm {Tr}}\big(\Cov\left( \Xin^{n-1}\right)\big) \le n \PCst  }
 h\left( \Yi_G^{n-1}| \Xin_{-\Mem}^{-1} = {\bf 0}_{\nt \cdot \Mem}\right) - h\left(\Wi_G^{n-1} \right)
\stackrel{(b)}{\! = \!} \CapGauss,
\label{eqn:AppBeq2}
\vspace{-0.2cm}
\end{align}
where ${\rm {Tr}}(\cdot)$ denotes the trace of a matrix, $(a)$ follows from \cite[Thm. 8.6.5]{Cover:06},
and since the differential entropy of a Gaussian random vector depends only on its covariance matrix \cite[Thm. 8.4.1]{Cover:06}, hence the supremum is carried out over the covariance of the input;
and $(b)$ follows from \cite[Lemma 3]{Brandenburg:74}, noting that $h\left( \Yi_G^{n-1}| \Xin_{-\Mem}^{-1} = {\bf 0}_{\nt \cdot \Mem}\right) - h\left(\Wi_G^{n-1} \right)$ denotes the mutual information between the input and the output of an \ac{lti} \ac{mimo} channel with additive Gaussian noise $\Wi_G^{n-1}$ and Gaussian output $\Yi_G^{n-1} \!=\! \tilde{\Hi}_n\Xin^{n-1} \!+\! \Wi^{n-1}$, as in \eqref{eqn:EqRelation}.
Plugging \eqref{eqn:AppBeq1}--\eqref{eqn:AppBeq2} into \eqref{eqn:AsymCapEq} yields
\vspace{-0.2cm}
\begin{align}
\Capacity
&< \CapGauss + \mathop{\lim}\limits_{n \rightarrow \infty} \left( \frac{1}{n}h\left(\Wi_G^{n-1} \right)- \frac{1}{n}h\left(\Wi^{n-1} \right) \right)
\! = \! \CapGauss + \HWiG -  \HWi,
\label{eqn:AppBeq3}
\vspace{-0.2cm}
\end{align}
which proves the upper bound in \eqref{eqn:UpperBoundCap1}.
\qed

\vspace{-0.2cm}
\subsection{Proof of Proposition \ref{pro:LowerBoundCap}}
\label{app:Proof4}
\vspace{-0.1cm}
The bound in \eqref{eqn:LowerBoundCap1a} follows since it can be concluded from \cite{DiggaviCover:2001}, \cite[Thm. 7.4.3]{Gallager:68}\footnote{While \cite[Thm. 7.4.3]{Gallager:68} is stated for scalar channels, the same proof also applies to \ac{mimo} channels.}, that for a given noise covariance matrix, then Gaussian noise is the worst-case noise distribution in terms of capacity, i.e., it results in the smallest capacity.  Specifically,  the supremum of $I\big(\Xin^{n-1} ; \Yi^{n-1} | \Xin_{-\Mem}^{-1} = {\bf 0}_{\nt \cdot \Mem} \big)= I\big(\Xin^{n-1} ; \tilde{\Hi}_n\Xin^{n-1} + \Wi^{n-1} \big)$ over all input distributions is lower bounded by the mutual information between the channel inputs and the channel outputs in which the additive non-Gaussian noise is replaced with an additive Gaussian noise with the same second-order moments as that of the non-Gaussian noise. Consequently, in the limit of $n \rightarrow \infty$, Eq. \eqref{eqn:LowerBoundCap1a} directly follows from \eqref{eqn:AsymCapEq}.

Next, from \eqref{eqn:EqRelation} we note that since both $\Xin^{n-1}$ and $\Wi^{n-1}$ are independent of $\Xin_{-\Mem}^{-1}$, then
\vspace{-0.2cm}
\begin{align}
h\left(\Yi^{n\!-\!1} | \Xin_{-\Mem}^{-1} \!=\! {\bf 0}_{\nt \cdot \Mem}\right)
&\! = \! h\left(\tilde{\Hi}_n\Xin^{n\!-\!1} \!+\! \Wi^{n\!-\!1} \right)
\stackrel{(a)}{\geq} \frac{n\cdot \nr}{2}\log\left(2^{\frac{2h\left( \tilde{\Hi}_n\Xin^{n\!-\!1}\right) }{n\cdot \nr}}\! +\! 2^{\frac{2h\left( \Wi^{n\!-\!1}\right) }{n\cdot \nr}}\right),
\label{eqn:AppCeq1}
\vspace{-0.2cm}
\end{align}
where $(a)$ follows from the entropy power inequality \cite[Thm. 17.7.3]{Cover:06}.
Thus, we have that
\vspace{-0.2cm}
\begin{align}
 &\mathop{\sup}\limits_{\Pdf{\Xin^{n-1}}:\; \frac{1}{n}\!\sum\limits_{i\! = \!0}^{n-1}\E \left\{\left\| \Xin\left[ i \right] \right\|^2\right\} \le \PCst  } \frac{1}{n}I\left(\Xin^{n-1} ; \Yi^{n-1}| \Xin_{-\Mem}^{-1} = {\bf 0}_{\nt \cdot \Mem} \right) \notag \\
	 &\quad\! = \! \mathop{\sup}\limits_{\Pdf{\Xin^{n-1}}:\; \frac{1}{n}\E \left\{\left\| \Xin^{n-1} \right\|^2\right\}\le \PCst  } \frac{1}{n}h\left(\Yi^{n-1}| \Xin_{-\Mem}^{-1} = {\bf 0}_{\nt \cdot \Mem} \right)  - \frac{1}{n}h\left(\Wi^{n-1} \right)  \notag \\
	 &\quad \stackrel{(a)}{\ge} \mathop{\sup}\limits_{\Pdf{\Xin^{n-1}}:\; \frac{1}{n}\E \left\{\left\| \Xin^{n-1} \right\|^2\right\}\le \PCst  } \frac{\nr}{2}\log\left(2^{\frac{2h\left( \tilde{\Hi}_n\Xin^{n-1}\right) }{n\cdot \nr}} + 2^{\frac{2h\left( \Wi^{n-1}\right) }{n\cdot \nr}}\right)  - \frac{1}{n}h\left(\Wi^{n-1} \right),
%
\label{eqn:AppCeq1a}
\vspace{-0.2cm}
\end{align}
where $(a)$ follows from \eqref{eqn:AppCeq1}.
Note that for any positive constants $a_1, a_2, a_3$ and a real constant $t$, the function $\log\left(a_1 2^{a_2 t} + a_3 \right)$ is monotonically increasing w.r.t. $t$, therefore
\vspace{-0.2cm}
\begin{align}
&\mathop{\sup}\limits_{\Pdf{\Xin^{n-1}}: \frac{1}{n}\E \left\{\left\| \Xin^{n-1} \right\|^2\right\}\le \PCst  } \frac{\nr}{2}\log\left(2^{\frac{2 }{n\cdot \nr}h\left( \tilde{\Hi}_n\Xin^{n-1}\right)} + 2^{\frac{2 }{n\cdot \nr}h\left( \Wi^{n-1}\right)}\right) \notag \\
&\qquad \qquad \! = \!
\frac{\nr}{2}\log\left(2^{\mathop{\sup}\limits_{\Pdf{\Xin^{n-1}}: \frac{1}{n}\E \left\{\left\| \Xin^{n-1} \right\|^2\right\}\le \PCst }\frac{2 }{n\cdot \nr}h\left( \tilde{\Hi}_n\Xin^{n-1}\right)} + 2^{\frac{2 }{n\cdot \nr}h\left( \Wi^{n-1}\right)}\right).
\label{eqn:AppCeq1b}
\vspace{-0.2cm}
\end{align}
Next, consider Eq. \eqref{eqn:AppCeq1b}: Note that when $\nt \! = \! \nr$ and $\Hi[0]$ is invertible, it follows from \eqref{eqn:DefHTilde} that $\tilde{\Hi}_n$ is also invertible, hence, by letting $\mathcal{M}_{n \cdot \PCst}$ be the set of $\nt \times \nt$ positive semi-definite real symmetric matrices $\myMat{C}_{\Xin}$ such that ${\rm Tr}\left( \myMat{C}_{\Xin}\right) \le n \cdot \PCst$, we have that
\vspace{-0.2cm}
\begin{align}
&\mathop{\sup}\limits_{\Pdf{\Xin^{n\!-\!1}}: \frac{1}{n}\E \left\{\left\| \Xin^{n\!-\!1} \right\|^2\right\}\le \PCst  }\frac{2} {n\cdot \nr} h\left( \tilde{\Hi}_n\Xin^{n\!-\!1}\right)
\stackrel{(a)}{\! = \!} \frac{2} {n\cdot \nr}\log|\tilde{\Hi}_n| + \frac{2} {n\cdot \nr}\mathop{\sup}\limits_{\Pdf{\Xin^{n\!-\!1}}: \frac{1}{n}\E \left\{\left\| \Xin^{n\!-\!1} \right\|^2\right\}\le \PCst  }\!\!\!\! h\left( \Xin^{n\!-\!1}\right) \notag \\
&\qquad\stackrel{(b)}{\! = \!} \frac{1} {n\cdot \nr}\log|\tilde{\Hi}_n|^2 + \frac{1} {n\cdot \nr}\mathop{\sup}\limits_{\Cov\left( \Xin^{n\!-\!1}\right) \in \mathcal{M}_{n \cdot \PCst} }\log\left( 2\pi e\right)^{n\cdot \nr}\left|\Cov\left( \Xin^{n\!-\!1}\right)\right|  \notag \\
&\qquad\! = \! \frac{1} {n\cdot \nr}\log|\tilde{\Hi}_n\tilde{\Hi}_n^T| +\log\left( 2\pi e\right) + \frac{1} {n\cdot \nr}\mathop{\sup}\limits_{\Cov\left( \Xin^{n\!-\!1}\right) \in \mathcal{M}_{n \cdot \PCst}  }\log\left|\Cov\left( \Xin^{n\!-\!1}\right)\right|,
\label{eqn:AppCeq5}
\vspace{-0.2cm}
\end{align}
where $(a)$ follows from \cite[Eq. (8.71)]{Cover:06}, and $(b)$ follows from \cite[Thm. 8.6.5]{Cover:06}. Since  $\Cov\left( \Xin^{n-1}\right)$ is positive semi-definite, it follows from the inequality of the arithmetic and geometric means \cite[Pg. 326]{Amann:05} that $\left|\Cov\left( \Xin^{n-1}\right)\right|  \le \left(\frac{1}{n\cdot \nt}{\rm Tr}\Big( \Cov\left( \Xin^{n-1}\right)\Big) \right)^{n\cdot \nt}$, and thus
$\frac{1}{n\cdot \nt} \log\left|\Cov\left( \Xin^{n-1}\right)\right|
 \! \le \! 
\log\left(\frac{1}{n\cdot \nt}{\rm Tr}\Big( \Cov\left( \Xin^{n-1}\right)\Big) \right)$.
 Consequently, 
\vspace{-0.2cm}
\begin{align}
\frac{1} {n\cdot \nt}\mathop{\sup}\limits_{\Cov\left( \Xin^{n-1}\right)\ \in \mathcal{M}_{n \cdot \PCst}  }\log\left|\Cov\left( \Xin^{n-1}\right)\right|
&\leq \mathop{\sup}\limits_{\Cov\left( \Xin^{n-1}\right) \in \mathcal{M}_{n \cdot \PCst} }\log\left(\frac{1}{n\cdot \nt}{\rm Tr}\Big( \Cov\left( \Xin^{n-1}\right)\Big) \right) \notag \\
&\stackrel{(a)}{\leq}  \log\left(\frac{\PCst}{\nt} \right),
\label{eqn:AppCeq7}
\vspace{-0.2cm}
\end{align}
where $(a)$ follows since $\log(\cdot)$ is monotonically increasing over $\mathcal{R}^+$. Note that for $\Cov\left( \Xin^{n-1}\right) \! = \! \frac{P}{\nt} \cdot \myMat{I}_{n \cdot \nt}$ the right hand side of \eqref{eqn:AppCeq7} is obtained with equality. Plugging this assignment into \eqref{eqn:AppCeq5}, and recalling that $\nt \! = \! \nr$, yields
\vspace{-0.2cm}
\begin{align}
\mathop{\sup}\limits_{\Pdf{\Xin^{n-1}}: \frac{1}{n}\E \left\{\left\| \Xin^{n-1} \right\|^2\right\}\le \PCst  }\frac{2} {n\cdot \nr} h\left( \tilde{\Hi}_n\Xin^{n-1}\right)
&\! = \!  \frac{1} {n\cdot \nr}\log|\tilde{\Hi}_n\tilde{\Hi}_n^T| +\log\left( 2\pi e\right) +  \log\left(\frac{\PCst}{\nt} \right) \notag \\
&\! = \! \log\left( 2\pi e\frac{\PCst}{\nt}\right) + \frac{1} {n\cdot \nr}\log|\tilde{\Hi}_n\tilde{\Hi}_n^T|.
\label{eqn:AppCeq8}
\vspace{-0.2cm}
\end{align}
Combining \eqref{eqn:AppCeq8}, \eqref{eqn:AppCeq1b}, and \eqref{eqn:AppCeq1a} results in
$ \frac{1}{n}
 I\left(\Xin^{n\!-\!1} ; \Yi^{n\!-\!1}| \Xin_{-\Mem}^{-1} \!= \!{\bf 0}_{\nt \cdot \Mem} \right)
 \!\ge\!  \frac{\nr}{2}\log\bigg(\frac{2\pi e \PCst}{\nt} \cdot 2^{\frac{1} {n\cdot \nr}\log|\tilde{\Hi}_n\tilde{\Hi}_n^T|} \!+ \!2^{\frac{2 }{n\cdot \nr}h\left( \Wi^{n\!-\!1}\right)}\bigg)  - \frac{1}{n}h\left(\Wi^{n\!-\!1} \right)$,
for any input distribution satisfying $\frac{1}{n}\E \left\{\left\| \Xin\left[ i \right] \right\|^2\right\} \le \PCst $ and for any $n$.
Lastly, we note that in the limit as $n\rightarrow \infty$, it follows from the extension of Szego's theorem to block-Toeplitz matrices \cite[Appendix A.2]{Brandenburg:74}, \cite[Thm. 5]{Guttierez:08} that
$\mathop{\lim}\limits_{n \rightarrow \infty} \frac{1}{n}\log\left|\tilde{\Hi}_n\tilde{\Hi}_n^T \right|
\! = \! \frac{1}{2\pi}\sum\limits_{k\! = \!0}^{ \nt -1} \int\limits_{\omega \! = \! -\pi}^{\pi} \log\left( \TlambdaN _k (\omega)\right)  d\omega$,
therefore, since $2^t$ is continuous w.r.t. $t \in \mathcal{R}$, letting $n$ tend to infinity in  \eqref{eqn:AppCeq1a}, it follows from  \eqref{eqn:AsymCapEq} and \cite[Pg. 224]{Amann:05} that
\vspace{-0.2cm}
\begin{eqnarray}
\Capacity 
   & \geq & \mathop{\lim}\limits_{n \rightarrow \infty} \frac{\nr}{2} \log\left(\frac{2\pi e \PCst}{\nt} \cdot  2^{\frac{1} {n\cdot \nr}\log|\tilde{\Hi}_n\tilde{\Hi}_n^T|}  +  2^{\frac{2 }{n\cdot \nr}h\left( \Wi^{n - 1}\right)}\right)  - \frac{1}{n}h\left(\Wi^{n - 1} \right)\nonumber\\
  & = & \frac{\nr}{2} \log\left(\frac{2\pi e \PCst}{\nt} \cdot 2^{\frac{1}{2\pi \cdot \nr}\sum\limits_{k\! = \!0}^{ \nt -1} \int\limits_{\omega \! = \! -\pi}^{\pi} \log\left( \TlambdaN _k (\omega)\right)  d\omega} + 2^{\frac{2}{\nr}\HWi}\right) - \HWi,
\label{eqn:AppCeq3}
\vspace{-0.2cm}
\end{eqnarray}
which completes the proof of \eqref{eqn:LowerBoundCap1}.
\qed

\vspace{-0.2cm}
\subsection{Proof of Theorem \ref{thm:MainResult1}}
\label{app:Proof2}
\vspace{-0.1cm}
The outline of the proof is as follows:
First, in Lemma \ref{Lem:Blocksize1} we show that the capacity of the MIMO \ac{bb}-\ac{plc} channel  \eqref{eqn:RxModel_1}, can be characterized by considering only codes whose blocklength is an integer multiple of $\Per$.
Then, we show that the capacity of MIMO \ac{bb}-\ac{plc} channels constrained to using only codes whose blocklength is an integer multiple of $\Per$ satisfies \eqref{eqn:MainResult1}.

\begin{lemma}
	\label{Lem:Blocksize1}
	The capacity of the MIMO \ac{bb}-\ac{plc} channel is identical to the maximum achievable rate obtained by considering only codes whose blocklength is an integer multiple of  $\Per$.
\end{lemma}
%
\begin{IEEEproof}
	The proof follows by first showing that any rate achievable for the MIMO \ac{bb}-\ac{plc} channel can be achieved by considering only codes whose blocklength is an integer multiple of $\Per$, and then showing   any rate achievable for the MIMO \ac{bb}-\ac{plc} channel when considering such codes, is an achievable rate for the MIMO \ac{bb}-\ac{plc} channel. As these steps are essentially the same as in the proof of   \cite[Lemma 1]{Shlezinger:16b}, they are not repeated here.
\end{IEEEproof}

Next, we note that the MIMO \ac{bb}-\ac{plc} channel  \eqref{eqn:RxModel_1} subject to the constraint that only codes whose blocklength is an integer multiple of $\Per$ are used, i.e.,  $\lequ \! = \! l \cdot \Per$ where $l \! \in \! \mathcal{N}$, can be represented  as an equivalent  $\Per  \times \Per$ \ac{lgmc} with code blocklength $l$ via the following assignments:
	Let the $\Per\cdot \tnt \times 1$ vector $\Xin\BB\left[i\right] \! \triangleq \! \Xeq_{i \cdot \Per}^{\left(i \! + \! 1\right)\cdot \Per\! -\! 1}$  be the input to the transformed channel and the  $\Per\cdot \tnt  \times 1$ vector $\Yi\BB\left[i\right] \! \triangleq \! \Yeq_{i \cdot \Per}^{\left(i \! + \! 1\right)\cdot \Per \! - \! 1}$ be the output of the channel. The transformation is clearly bijective as  for the \ac{bb}-\ac{plc} channel we consider only codes whose blocklength is an integer multiple of $\Per$.
	For each blocklength $l$, the input to the equivalent \ac{lgmc} satisfies
	\vspace{-0.2cm}
	\begin{equation*}
	\frac{1}{l}\sum\limits_{i  =  0}^{l - 1} \E\left\{\left\| \Xin\BB\left[ i \right] \right\|^2\right\}
	\! = \! \frac{1}{l}\sum\limits_{i \! = \! 0}^{l - 1}\sum\limits_{k = 0}^{\Per - 1} \E\left\{\left\| \Xeq\left[ i\cdot \Per + k \right] \right\|^2\right\}
	\! = \! \frac{\Per}{\lequ}\sum\limits_{\ieq  =  0}^{\lequ - 1} \E\left\{\left\| \Xeq\left[
\;\ieq \;\right] \right\|^2 \right\}
	\stackrel{(a)}{\leq} \Per \cdot \PCsteq,
	\vspace{-0.3cm}
	\end{equation*}
	where $(a)$ follows from \eqref{eqn:Constraint0}.
	Consequently, the equivalent \ac{lgmc} input is subject to a maximal power constraint $\PCst\BB \! = \!\Per \cdot \PCsteq$.
	Next, we note that the input-output relationship of the \ac{bb}-\ac{plc} channel \eqref{eqn:RxModel_1} implies that the input-output relationship of the transformed channel is given by \eqref{eqn:RxModel_3}, and  that the equivalent  \ac{lgmc} noise $\Wi\BB\left[i\right]$ appearing in \eqref{eqn:RxModel_3}, is a zero-mean strict-sense stationary process. Moreover, as $\Per > \Memeq$, it follows that the temporal dependence of $\Wi\BB\left[i\right]$ spans an interval of length $\Mem \! = \! 1$. Recall that $\TCapacity$ denotes the capacity of the channel \eqref{eqn:RxModel_3}--\eqref{eqn:ConstraintBB}.
	
	As each channel use in the equivalent \ac{lgmc} \eqref{eqn:RxModel_3}--\eqref{eqn:ConstraintBB} corresponds to $\Per$ channel uses in the \ac{bb}-\ac{plc} channel \eqref{eqn:RxModel_1}--\eqref{eqn:Constraint0}, it follows that the maximal achievable rate of the  \ac{bb}-\ac{plc} channel, measured in bits per channel use, subject to the restriction that only codes whose blocklength is an integer multiple of $\Per$ are allowed, can be obtained from the maximal achievable rate of the equivalent \ac{lgmc} as $ \Capeq\! = \! \frac{1}{\Per}\TCapacity$.
	Finally, from Lemma \ref{Lem:Blocksize1}, we conclude that $\Capeq$ is the maximum achievable rate for the \ac{bb}-\ac{plc} channel, thus proving the theorem.
\qed

\vspace{-0.2cm}
\subsection{Proof of Proposition \ref{thm:NakagamiEntropy}}
\label{app:NakagamiEntropy}
\vspace{-0.1cm}
In order to derive the differential entropy of complex Nakagami-$m$ RVs, we
use the
following lemma, which states the \ac{pdf} of a family of complex \acp{rv}:
\begin{lemma}
	\label{lem:EntComplexRV}
	Let $\WW$ be a complex RV given by $\WW \! = \! Xe^{j\Theta}$, where $X$ is a non-negative real RV, and $\Theta$ is an \ac{rv} uniformly distributed over $[0,2\pi]$, mutually independent of $X$, then, the \ac{pdf} of $\WW$ is given by
$	\pdf{\WW}{\ww}\! = \! \frac{\pdf{X}{\left| \ww\right| }}{2\pi\left|\ww\right|}$,
	and its differential entropy is given by
	\vspace{-0.1cm}
	\begin{equation}
		\label{eqn:EntComplexRV2}
	h(\WW)\! = \! \log(2\pi)+\E\Big\lbrace\log(X)\Big\rbrace+h(X).
		\vspace{-0.1cm}
	\end{equation}
\end{lemma}
\begin{IEEEproof}
Let $\WWr, \WWi$ be the real and imaginary parts of $W$, respectively, and recall that the \ac{pdf} of a complex RV $\WW = \WWr +j\WWi$ is given by  $\pdf{\WW}{\ww\! = \!\wwr+j\wwi}\! = \!\pdf{\WWr,\WWi}{\wwr,\wwi}$ \cite[Pg. 188]{Papoulis:91}.
Consequently,  letting $\arg (z)$ denote the phase of a complex number $z$, the \ac{pdf} $\pdf{\WWr,\WWi}{\wwr,\wwi}$ is obtained using the transformation of RVs theorem  as in \cite[Pg. 146]{Papoulis:91}:
\vspace{-0.2cm}
\begin{equation}
\pdf{\WWr,\WWi}{\wwr,\wwi}
\! = \!\frac{\pdf{X,\Theta}{\sqrt{\wwr^2+\wwi^2},\arg\left(\frac{\wwi}{\wwr} \right) }}{\sqrt{\wwr^2+\wwi^2}} 
\stackrel{(a)}{\! = \!}\frac{\pdf{X}{\sqrt{\wwr^2+\wwi^2}}}{2\pi\sqrt{\wwr^2+\wwi^2}} \! = \!\frac{\pdf{X}{\left| \ww\right| }}{2\pi\left|\ww\right|},
\label{eqn:CNekagaemyPDF1}
\vspace{-0.2cm}
\end{equation}
where $(a)$ follows since $X$ and $\Theta$ are mutually independent, thus $\pdf{X,\Theta}{x,\theta}\! = \!\pdf{X}{x}\pdf{\Theta}{\theta}$, and from the uniform distribution of $\Theta$. It thus follows that $	\pdf{\WW}{\ww}\! = \! \frac{\pdf{X}{\left| \ww\right| }}{2\pi\left|\ww\right|}$.

Using \eqref{eqn:CNekagaemyPDF1}, we next derive the differential entropy of $\WW$ as:
\vspace{-0.2cm}
\begin{align}
h(\WW)
&\! = \!-\int\limits_{\mathcal{R}^2}\frac{\pdf{X}{\sqrt{\wwr^2+\wwi^2}}}{2\pi\sqrt{\wwr^2+\wwi^2}}\log\left(\frac{\pdf{X}{\sqrt{\wwr^2+\wwi^2}}}{2\pi\sqrt{\wwr^2+\wwi^2}} \right)d\wwr d\wwi \notag \\
&\stackrel{(a)}{\! = \!}-\int\limits_{\theta\! = \!0}^{2\pi}\int\limits_{x\! = \!0}^{\infty}x\frac{\pdf{X}{x}}{2\pi x}\log\left(\frac{\pdf{X}{x}}{2\pi x} \right)dx d\theta
\! = \! -\int\limits_{x\! = \!0}^{\infty}\pdf{X}{x}\log\left( \frac{\pdf{X}{x}}{2\pi x}\right)dx,
\label{eqn:CNekagaemyEntropy1}
\vspace{-0.2cm}
\end{align}
where $(a)$ is obtained by switching the integration variables from $(\wwr,\wwi)$ to $(x, \theta)$, given by $x\! = \!\sqrt{\wwr^2+\wwi^2}$ and $\theta\! = \!\tan^{-1}\left(\frac{\wwi}{\wwr}\right)$. Note that \eqref{eqn:CNekagaemyEntropy1} can be written as
\vspace{-0.2cm}
\begin{align}
-\!\int\limits_{x\! = \!0}^{\infty}\pdf{X}{x}\!\log\left( \frac{\pdf{X}{x}}{2\pi x}\right)\! dx
&\! = \!  \int\limits_{x\! = \!0}^{\infty}\!\pdf{X}{x}\log(2\pi)dx \!+\! \int\limits_{x\! = \!0}^{\infty}\!\pdf{X}{x}\log(x)d x\!-\! \int\limits_{x\! = \!0}^{\infty}\!\pdf{X}{x}\log\left(\pdf{X}{x}\right)dx \notag \\
& \! = \! \log(2\pi)+\E\left\{\log(X)\right\}+h(X).
\label{eqn:CNekagaemyEntropy1a}
\vspace{-0.2cm}
\end{align}
Plugging \eqref{eqn:CNekagaemyEntropy1a} into  \eqref{eqn:CNekagaemyEntropy1} we obtain \eqref{eqn:EntComplexRV2}.
\end{IEEEproof}

For a complex Nakagami-$m$ RV $\WW$, we have that the \ac{pdf} of $X$  is given by \eqref{eqn:RNekagaemyPDF2}.
Plugging the \ac{pdf} \eqref{eqn:RNekagaemyPDF2} into  \eqref{eqn:CNekagaemyPDF1} we obtain the \ac{pdf} of $\WW$ as:
$\pdf{\WW}{\ww}\! = \!\frac{2}{2\pi\cdot \Gamma\left(m\right)}\left( \frac{m}{\Omega}\right)^m\left| \ww\right|^{2m-2}e^{-\frac{m\left| \ww\right|^2}{\Omega}}$.
%
%
To obtain the differential entropy of the complex Nakagami-$m$ RV $\WW = Xe^{j\Theta}$, we note that for  $X \sim \Nakagami{m}{\Omega}$,
$\E\left\{ \log(X)\right\}
\! = \!\int\limits_{x\!=\!0}^{\infty}\frac{2}{\Gamma(m)}\left(\frac{m}{\Omega} \right)^m x^{2m-1}e^{-\frac{mx^2}{\Omega}}\log(x)dx$.
Setting $t\triangleq \frac{mx^2}{\Omega}$ as the integration variable, 
we have $dt\! = \!2\frac{mx}{\Omega}dx$, $\log(t)\! = \log\left( \frac{m}{\Omega}\right)+2\log(x)$, and $x^2\! = \!\frac{\Omega}{m}t$, resulting in:
\vspace{-0.2cm}
\begin{align}
\E\left\{\log(X)\right\} &\! = \!
\int\limits_{t\! = \!0}^{\infty}\frac{1}{2\Gamma(m)}
t^{m-1}e^{-t}\bigg(\log(t)-\log\left( \frac{m}{\Omega}\right) \bigg)dt \notag\\
&\! = \!\frac{1}{2\ln(2)}\int\limits_{t\! = \!0}^{\infty}\frac{1}{\Gamma(m)}t^{m-1}e^{-t}\ln(t)dt-\frac{1}{2\Gamma(m)}\log\left( \frac{m}{\Omega}\right)\int\limits_{t\! = \!0}^{\infty}t^{m-1}e^{-t}dt \notag \\
&\stackrel{(a)}{\! = \!}\frac{1}{2\ln(2)}\Psi(m)-\frac{1}{2}\log\left( \frac{m}{\Omega}\right),
\label{eqn:CNekagaemyEntropy4}
\vspace{-0.2cm}
\end{align}
where $(a)$ follows since  $\Psi(x)\! = \!\frac{d}{dx}\Big( \ln\big(\Gamma(x) \big) \Big)\! = \!\frac{1}{\Gamma(x)}\int\limits_{t\! = \!0}^{\infty}t^{x-1}e^{-t}\ln(t)dt$ \cite[Tbl. 0.1]{Michalowicz:13}.
Next, recall that the differential entropy of a real-valued Nakagami-$m$ RV is given by \cite[Ch. 4.18]{Michalowicz:13}:
$h(X)\! = \!\log\left(\frac{\Gamma(m)}{2}\sqrt{\frac{\Omega}{m}}e^{\frac{2m-(2m-1)\Psi(m)}{2}} \right)$.
Plugging this and \eqref{eqn:CNekagaemyEntropy4} 
into \eqref{eqn:EntComplexRV2}, we have that
\vspace{-0.2cm}
\begin{align}
\label{eqn:CNekagaemyEntropy6}
h\left(\WW \right)
= h(\WWr, \WWi)
& \! = \! \log(2\pi) \! + \! \frac{1}{2\ln(2)}\Psi(m)-\frac{1}{2}\log\left( \frac{m}{\Omega}\right)+\log\left(\frac{\Gamma(m)}{2}\sqrt{\frac{\Omega}{m}}e^{\frac{2m-(2m-1)\Psi(m)}{2}} \right) \notag \\
&\! = \! \frac{1}{2\ln(2)}\Psi(m)+ \log\left(\frac{\pi\Omega}{m}\Gamma(m)e^{\frac{2m-(2m-1)\Psi(m)}{2}} \right),
\vspace{-0.2cm}
\end{align}
proving the proposition.
\qed

\vspace{-0.2cm}
\subsection{Proof of Lemma \ref{cor:entropyGain}}
\label{app:EntropyGain}
\vspace{-0.1cm}
To prove Lemma \ref{cor:entropyGain}, we first state Lemma \ref{lem:entropyGain}, which characterizes a relationship between the entropy rate of an i.i.d. process $\Ui[i]$, $\HUi \!=\! h\big( \Ui\big),$  and the entropy rate of $\Wi\left[i\right]$, $\HWi$, obtained by \ac{lti} filtering of $\Ui[i]$:
\begin{lemma}
	\label{lem:entropyGain}
	Let $\Ui[i] \in \mathcal{R}^{\nr}$ be an i.i.d. multivariate process, $\{\Fi[\tau] \}_{\tau = 0}^{\Mem}$ be a set of $\nr\times \nr$ matrices s.t. $\Fi[0]$ is non-singular.
	Define  $\Wi[i] = \sum\limits_{\tau=0}^{\Mem} \Fi[\tau] \Ui[i-\tau]$, and $\Ffd(\omega) \!\triangleq \! \sum\limits_{\tau=0}^\Mem\!\Fi[\tau]e^{-j\omega \tau}$, and let  $\HWi$ and  $\HUi$ denote the entropy rates of $\Wi[i]$ and $\Ui[i]$,  respectively. Then, we have
	\vspace{-0.2cm}
	\begin{equation}
	\label{eqn:EntropyGaina}
	\HWi=\frac{1}{2\pi}\int\limits_{\omega=0}^{2\pi}\log \left| \Ffd\left( \omega\right) \right| d\omega+\HUi.
	\vspace{-0.2cm}
	\end{equation}
\end{lemma}
\begin{remark}
	For $\nr \!=\! 1$,  \eqref{eqn:EntropyGaina} specializes the entropy gain of scalar filters in \cite[Thm. 14]{Shannon:48}.
\end{remark}

\begin{IEEEproof}
Since we are interested in the entropy rate we may assume that the blocklengths are sufficiently large and consider $n > 2 \Mem$. Define the  $n\cdot \nr \times n \cdot \nr$ matrix $\tilde{\Fi}_n^a$, the  $\Mem\cdot \nr \times \Mem \cdot \nr$ matrix $\tilde{\Fi}_\Mem^b $, and the  $n\cdot \nr \times \Mem \cdot \nr$ matrix $\tilde{\Fi}_n^c $, via
\vspace{-0.2cm}	
\begin{equation}
\label{eqn:DefFTilde}
\tilde{\Fi}_n^a \!\triangleq\! \left[ {\begin{array}{*{20}{c}}
	{\Fi\!\left[0\right]}&\cdots &0&  \cdots &0\\
	\vdots & \ddots &{}&\ddots& \vdots \\
	{\Fi\!\left[\Mem\right]}& \cdots &{\Fi\!\left[0\right]}& \cdots &0\\
	\vdots & \ddots &{}& \ddots & \vdots\\
	0& \cdots &{\Fi\!\left[\Mem\right]}& \cdots &{\Fi\!\left[0\right]}
	\end{array}} \right],\quad
\tilde{\Fi}_\Mem^b \!\triangleq\! \left[ {\begin{array}{*{20}{c}}
	{\Fi\!\left[\Mem\right]}&\cdots &{\Fi\!\left[1\right]}\\
	\vdots & \ddots & \vdots \\
	0& \cdots &{\Fi\!\left[\Mem\right]}&
	\end{array}} \right],\quad
\tilde{\Fi}_n^c \!\triangleq\! \left[ {\begin{array}{*{20}{c}}
	\tilde{\Fi}_\Mem^b\\
	\myMat{0}_{(n-\Mem)\cdot \nr \times \Mem \cdot \nr}
	\end{array}} \right].
\vspace{-0.2cm}
\end{equation}
Note that $\tilde{\Fi}_n^a$ is block-Toeplitz and non-singular (hence, invertible), as $\Fi[0]$ is non-singular.
Using \eqref{eqn:DefFTilde}, we can write $\Wi^{n-1} \! = \! \tilde{\Fi}_n^a\Ui^{n-1} + \tilde{\Fi}_n^c\Ui_{-\Mem}^{-1} = \tilde{\Fi}_n^e \Ui^{n-1}_{-m}$. As $\Ui[i]$ is i.i.d., then  $\tilde{\Fi}_n^a\Ui^{n-1}$ and $\tilde{\Fi}_n^c\Ui_{-\Mem}^{-1}$ are mutually independent. Hence, $h\left(\Wi^{n-1}\big| \tilde{\Fi}_n^c\Ui_{-\Mem}^{-1} \right) \! = \!  h\left(\tilde{\Fi}_n^a\Ui^{n-1} \right)$, and
we can write
\vspace{-0.2cm}
\begin{align}
&h\left(\Wi^{n-1} \right) - h\left(\tilde{\Fi}_n^a\Ui^{n-1} \right)
\! = \! I\left(\tilde{\Fi}_n^c\Ui_{-\Mem}^{-1}; \Wi^{n-1}  \right)
\stackrel{(a)}{\! = \!} I\left(\tilde{\Fi}_m^b\Ui_{-\Mem}^{-1}; \Wi^{n-1}  \right)
\label{eqn:DifEnt1} \\
&\quad =  I\left(\tilde{\Fi}_m^b\Ui_{-\Mem}^{-1}; \tilde{\Fi}_n^a\Ui^{n-1} + \tilde{\Fi}_n^c\Ui_{-\Mem}^{-1}  \right)
=  h\left(\tilde{\Fi}_m^b\Ui_{-\Mem}^{-1}\right)- h\left(\tilde{\Fi}_m^b\Ui_{-\Mem}^{-1}| \tilde{\Fi}_n^a\Ui^{n-1} + \tilde{\Fi}_n^c\Ui_{-\Mem}^{-1} \right)\nonumber\\
&\quad \stackrel{(b)}{\le}  h\left(\tilde{\Fi}_m^b\Ui_{-\Mem}^{-1}\right)- h\left(\tilde{\Fi}_m^b\Ui_{-\Mem}^{-1}| \tilde{\Fi}_n^a\Ui^{n-1} + \tilde{\Fi}_n^c\Ui_{-\Mem}^{-1}, \Ui_m^{n-1} \right)\nonumber\\
&\quad \stackrel{(c)}{\! = \!}  h\left(\tilde{\Fi}_m^b\Ui_{-\Mem}^{-1}\right)- h\left(\tilde{\Fi}_m^b\Ui_{-\Mem}^{-1}| \tilde{\Fi}_{2m}^a\Ui^{2m-1} + \tilde{\Fi}_{2m}^c\Ui_{-\Mem}^{-1}, \Ui_{m}^{n-1} \right)\nonumber\\
&\quad \stackrel{(d)}{\! = \!}  h\left(\tilde{\Fi}_m^b\Ui_{-\Mem}^{-1}\right)- h\left(\tilde{\Fi}_m^b\Ui_{-\Mem}^{-1}| \tilde{\Fi}_{2m}^a\Ui^{2m-1} + \tilde{\Fi}_{2m}^c\Ui_{-\Mem}^{-1}, \Ui_{m}^{2m-1} \right)\nonumber\\
&\quad =  I\left(\tilde{\Fi}_m^b\Ui_{-\Mem}^{-1}; \tilde{\Fi}_{2m}^a\Ui^{2m-1} + \tilde{\Fi}_{2m}^c\Ui_{-\Mem}^{-1}, \Ui_{m}^{2m-1} \right)
=  I\left(\tilde{\Fi}_m^b\Ui_{-\Mem}^{-1}; \Wi^{2m-1}, \Ui_{m}^{2m-1} \right),
\vspace{-0.1cm}
\label{eqn:DifEnt11}
\end{align}
where $(a)$ follows from the definition of $\tilde{\Fi}_n^c$ in \eqref{eqn:DefFTilde}; and (b) follows as conditioning decreases the entropy; in $(c)$ the matrix $\tilde{\Fi}_{2\Mem}^a$ is an $2\Mem\cdot \nr \times 2\Mem\cdot
\nr$ matrix in which each row consists of the first $2\Mem$ elements of the corresponding row of $\tilde{\Fi}_n^a$, and $\tilde{\Fi}_{2m}^c$ is a matrix which consists of the first $2m$ rows of
$\tilde{\Fi}_n^c$. Lastly, $(d)$ follows as $\Ui\left[\,\ieq\,\right]$ is an i.i.d. sequence.
Noting that Eq. \eqref{eqn:DifEnt1} implies that $h\left(\Wi^{n-1} \right) \ge h\left(\tilde{\Fi}_n^a\Ui^{n-1} \right)$, we have that
\vspace{-0.1cm}
\begin{align}
0 \le h\left(\Wi^{n-1} \right) - h\left(\tilde{\Fi}_n^a\Ui^{n-1} \right)
&\stackrel{(a)}{\le}I\left(\tilde{\Fi}_m^b\Ui_{-\Mem}^{-1}; \Wi^{2m-1}, \Ui_{m}^{2m-1} \right),
\label{eqn:DifEnt2}
\vspace{-0.1cm}
\end{align}
where $(a)$ follows from \eqref{eqn:DifEnt11}.
%
%
%
%
Observing that the right hand side of \eqref{eqn:DifEnt2} is a finite value which does not depend on $n$, then, dividing both sides of \eqref{eqn:DifEnt1} by $n$ and letting $n$ tend to infinity yields $\mathop{\lim}\limits_{n\rightarrow \infty}\frac{1}{n}h\left(\Wi^{n-1} \right)  - \mathop{\lim}\limits_{n\rightarrow \infty}\frac{1}{n}h\big(\tilde{\Fi}_n^a\Ui^{n-1}\big) = 0$. Therefore,
\vspace{-0.1cm}
\begin{equation*}
\HWi
\! = \! \mathop{\lim}\limits_{n\rightarrow \infty}\frac{1}{n}h\left(\tilde{\Fi}_n^a\Ui^{n-1}\right)
\stackrel{(a)}{\! = \!} \mathop{\lim}\limits_{n\rightarrow \infty}\left( \frac{1}{n}
\log\left|\tilde{\Fi}_n^a\right|  +  \frac{1}{n}h\left(\Ui^{n-1}\right) \right)
\stackrel{(b)}{\! = \!}\frac{1}{2\pi}\int\limits_{\theta\! = \!0}^{2\pi}\log \left| \Ffd\left( \theta\right) \right|d\theta +\HUi,
\vspace{-0.1cm}
\end{equation*}
where $(a)$ follows from \cite[Eq. (8.71)]{Cover:06} as $\tilde{\Fi}_n^a$ is invertible, and $(b)$ follows from the extension of Szego's theorem to block-Toeplitz matrices \cite[Thm. 5]{Guttierez:08}.
\end{IEEEproof}
Since by \eqref{eqn:LPTVFilt2}, $\Wi\BB\left[\,\ieq\,\right] $ is the output of an \ac{lti} filter with i.i.d. input $\Ui\left[\,\ieq\,\right]$, and as that the entropy rate of $\Ui\left[\ieq \right]$ is given by $\Per \cdot h\big(\Ueq\big)$, it follows from Lemma \ref{lem:entropyGain} that $\HWiBB=\frac{1}{2\pi}\int\limits_{\omega=0}^{2\pi}\log \left| \Ffd\left( \omega\right) \right| d\omega + \Per \cdot h\left(\Ueq \right)$, proving the lemma.
\qed

\end{appendix}



\vspace{-0.2cm}
\begin{thebibliography}{10}
\vspace{-0.1cm}	
{\setstretch{1.2}
	\setlength{\itemsep}{-1pt}
	\bibitem{Ferreira:10}
	H. C. Ferreira, L. Lampe. J. Newbury, and T. G. Swart.
	\newblock{\em Power Line Communications - Theory and Applications for Narrowband and Broadband Communications over Power Lines}.
	\newblock Wiley and Sons, Ltd., 2010.

	\bibitem{Cano:16}
	C. Cano, A. Pittolo, D. Malone, L. Lampe, A. M. Tonello, and A. G. Dabak.
	\newblock{``State of the art in power line communications: From the applications to the medium,"}
	\newblock{\em IEEE J. Sel. A. Commun.}, vol. 34, no. 7, Jul. 2016, pp. 1935--1952.
	
	\bibitem{Berger:15}
	L. T. Berger, A. Schwager, P. Pagani, and D. M. Schneider.
	\newblock{``MIMO power line communications,"}
	\newblock{\em IEEE Commun. Surveys \& Tutorials}, vol. 17, no. 1, Q1 2015, pp. 106--124.	
	
	\bibitem{Meng:05}
	H. Meng, Y. L. Guan, and S. Chen.
	\newblock{``Modeling and analysis of noise effects on broadband power-line communications,"}
	\newblock{\em IEEE Trans. Power Del.}, vol. 20, no. 2, Apr. 2005, pp. 630 -- 637.	
	
	
%
	\bibitem{Mathur:14}
	A. Mathur and M. R. Bhatnagar.
	\newblock{``PLC performance analysis assuming BPSK modulation over Nakagami-m additive noise,"}
	\newblock{\em IEEE Commun. Lett.}, vol. 18, no. 6, Jun. 2014, pp. 909 -- 912.
	
	\bibitem{Mathur:15}
	A. Mathur, M. R. Bhatnagar, and B. K. Panigrahi.
	\newblock{``Performance evaluation of PLC Under the combined effect of background and impulsive noises,"}
	\newblock{\em IEEE Commun. Lett.}, vol. 19, no. 7, Jul. 2015, pp. 1117 -- 1120.	

	\bibitem{Bo:07}
	W. Bo, Q. Yinghao, H. Peiwei, and C. Wenhao.
	\newblock{``Indoor powerline channel simulation and capacity analysis,"}
	\newblock{\em IET Conference on Wireless, Mobile Sensor Networks}, Shanghai, China, Dec. 2007, pp. 154-156.

	\bibitem{Gianaroli:14}
	F. Gianaroli, F.. Pancaldi, and G. M. Vitetta.
	\newblock{``The impact of statistical noise modeling on the error-rate performance of OFDM power-line communications,"}
	\newblock{\em IEEE Trans. Power Del.}, vol. 29, no. 6, Apr. 2014, pp. 2622 -- 2630.
	
	\bibitem{Gotz:04}
	M. Gotz,  M. Rapp, and  K. Dostert.
	\newblock{``Power line channel characteristics and their effect on communication system design,"}
	\newblock{\em IEEE Commun. Mag.}, vol. 42, no. 4, Apr. 2004, pp. 78 -- 86.

	\bibitem{Tonello:14}
	A. M. Tonello, F. Versolatto, and A. Pittolo.
	\newblock{``In-home power line communication channel: Statistical characterization,"}
	\newblock{\em IEEE Trans. Commun.}, vol. 62, no. 6, Jun. 2014, pp. 2096 -- 2106.

	\bibitem{Esmailian:03}
	T. Esmailian, F. R. Kschischang, and P. Glenn Gulak.
	\newblock{``In-building power lines as high-speed communication channels: channel characterization and a test channel ensemble,"}
	\newblock{\em Int. J.  Commun. Sys.}, vol. 16, no. 5, May 2003, pp. 381-400.

	\bibitem{Galli:11}
	S. Galli.
	\newblock{``A novel approach to the statistical modeling of wireline channels,"}
	\newblock{\em IEEE Trans. Commun.}, vol. 59, no. 5, May 2011, pp. 1332-1345.


	\bibitem{Corripio:06}
	F. J. Ca{\~n}ete, J. A. Cort{\'e}s, L. D{\'\i}ez, and J. T.  Entrambasaguas.
	\newblock{``Analysis of the cyclic short-term variation of indoor power line channels,"}
	\newblock{\em IEEE J. Sel. A. Commun.}, vol. 24, no. 7, Jul. 2006, pp. 1327--1338.
	
	\bibitem{Cortes:10}
	J. A. Cort{\'e}s, L. D{\'\i}ez, F. J. Ca{\~n}ete, and J. J. S{\'a}nchez-Marti{\'n}ez,.
	\newblock{``Analysis of the indoor broadband power-line noise scenario,"}
	\newblock{\em IEEE Trans.  Electromagn. Compat.}, vol. 52, no. 4, Nov. 2010,  pp. 849--858.

	\bibitem{Zimmermann:02b}
	M. Zimmermann and K. Dostert.
	\newblock{``Analysis and modeling of impulsive noise in broad-band powerline communications,"}
	\newblock{\em IEEE Trans.  Electromagn. Compat.}, vol. 44, no. 1, Feb. 2002,  pp. 249--258.
	
	
	\bibitem{Ma:05}
	Y. H. Ma, P. L. So, and E. Gunawan.
	\newblock{``Performance analysis of OFDM systems for broadband power line communications under impulsive noise and multipath effects,"}
	\newblock{\em IEEE Trans.  Power Del.}, vol. 20, no. 2, Apr. 2005,  pp. 674--681.


	\bibitem{Zimmermann:02a}
	M. Zimmermann and K. Dostert.
	\newblock{``A multipath model for the powerline channel,"}
	\newblock{\em IEEE Trans.  Commun.}, vol. 50, no. 4, Apr. 2002,  pp. 553--559.
	
	\bibitem{Gianaroli:14a}
	F. Gianaroli, F. Pancaldi, and G. M. Vitetta.
	\newblock{``On the use of Zadeh's series expansion for modeling and estimation of indoor powerline channels,"}
	\newblock{\em IEEE Trans. Commun.}, vol. 62, no. 7, Jul. 2014, pp. 2558--2568.
		
	\bibitem{Corripio:11}
	F. J. Ca{\~n}ete, J. A. Cort{\'e}s,  L. D{\'\i}ez, and J. T. Entrambasaguas.
	\newblock{``A channel model proposal for indoor power line communications,"}
	\newblock{\em IEEE Commun. Mag.}, vol. 49, no. 12, Dec. 2011,  pp. 166--174.	
		
	
	\bibitem{Rende:11}
	D. Rende, A. Nayagam, K. Afkhamie, L. Yonge, R. Riva, D. Veronesi, F. Osnato, and P. Bisaglia.
	\newblock{``Noise correlation and its effect on capacity of inhome MIMO power line channels,"}
	\newblock{\em IEEE International Symposium on Power-Line Communications and its Applications (ISPLC)}, Udine, Italy, Apr. 2011, pp. 60--65.

	
	\bibitem{Pagani:16}
	P. Pagani and A. Schwager.
	\newblock{``A statistical model of the in-home MIMO PLC channel based on European field measurements,"}
	\newblock{\em IEEE J. Sel. A. Commun.}, vol. 34, no. 7, Jul. 2016, pp. 2033--2044.

	\bibitem{Vernosi:11}
	D. Veronesi, R. Riva, P. Bisaglia, F. Osnato, K. Afkhamie, A. Nayagam, D. Rende, and L. Yonge.
	\newblock{``Characterization of in-home MIMO power line channels,"}
	\newblock{\em IEEE International Symposium on Power-Line Communications and its Applications (ISPLC)}, Udine, Italy, Apr. 2011, pp. 42--47.
		
	\bibitem{Corchado:16}
	J. A. Corchado,  J. A. Cort{\'e}s, F. J. Ca{\~n}ete, A. Arregui, and L. D{\'\i}ez.
	\newblock{``Analysis of the spatial correlation of indoor MIMO PLC channels,"}
	\newblock{\em IEEE Commun. Letters}, vol. 21, no. 1, Jan. 2017,  pp. 40--43.
	
	\bibitem{Middleton:77}
	D. Middleton.
	\newblock{``Statistical-physical models of electromagnetic interference,"}
	\newblock{\em IEEE Trans. Electromagn. Compat.}, vol. 19, no. 3, Aug. 1977, pp. 106 -- 127.

	\bibitem{Nassar:11}
	M. Nassar, K. Gulati, Y. Mortazavi, and B. L. Evans.
	\newblock{``Statistical modeling of asynchronous impulsive noise in powerline communication networks,"}
	\newblock{\em IEEE Global Communications Conference (GLOBECOM).}, Houston, TX, Dec. 2011.	
	
	\bibitem{Lampe:13}
	M. A. Tunc, E. Perrins, and L. Lampe.
	\newblock{``Optimal LPTV-aware bit loading in broadband PLC,"}
	\newblock{\em IEEE Trans. Commun.}, vol. 61, no. 12, Dec. 2013, pp. 5152--5162.
		
	\bibitem{Cover:06}		
	T. M. Cover and J. A. Thomas.
	\newblock{\em Elements of Information Theory}.
	\newblock Wiley Press, 2006.
	
	\bibitem{Le:16}
	D. A. Le, H. V. Vu, N. H. Tran, M. C. Gursoy, T. Le-Ngoc.
	\newblock{``Approximation of achievable rates in additive Gaussian mixture noise channels,"}
	\newblock{\em IEEE Trans. Commun.}, vol. 64, no. 23, Dec. 2016, pp. 5011 - 5024.

	
	\bibitem{Shlezinger:15}
	N. Shlezinger and R. Dabora.
	\newblock{``On the capacity of narrowband PLC channels,"}
	\newblock{\em IEEE Trans. Commun.}, vol. 63, no. 4, Apr. 2015, pp. 1191 - 1201.	

	
	\bibitem{Gallager:68}	
	R. G. Gallager.
	\newblock{\em Information Theory and Reliable Communication}.
	\newblock{\em } Wiley and Sons, Ltd., 1968.

	\bibitem{Goldsmith:01}
	A. Goldsmith and M. Effros.
	\newblock{``The capacity region of broadcast channels with intersymbol interference and colored Gaussian noise,"}
	\newblock{\em IEEE Trans. Inform. Theory}, vol. 47, no. 1, Jan. 2001, pp. 219--240.
	
	\bibitem{Weingerten:06}
	H. Weingerten, Y. Steinberg, and S. Shamai,
	\newblock{``The capacity region of the Gaussian multiple-input multiple-output broadcast channel,"}
	\newblock{\em IEEE Trans. Inform. Theory}, vol. 52, no. 9, Sep. 2006, pp. 3936--3964.

	\bibitem{Dobrushin:63}
	R. L. Dobrushin.
	\newblock{``General formulation of Shannon’s main theorem in information theory,"}
	\newblock{\em Amer. Math. Soc. Translations: Series 2}, vol. 33, 1963, pp. 323--438.	

    \bibitem{Tsaregradskii:58}
    I. P. Tsaregradskii.
    \newblock{``A note on the capacity of a stationary channel with finite memory,"}
    \newblock{\em Theory of Probability and its Applications}, vol. 3, no. 1, 1958, pp. 79--91.
	
	\bibitem{Han:03}
	T. S. Han.
	\newblock{\em Information-Spectrum Methods in Information Theory}.
	\newblock Springer, 2003.
		
	\bibitem{Massey:88}
	W. Hirt and J. L. Massey.
	\newblock{``Capacity of discrete-time Gaussian channel with intersymbol interference,"}
	\newblock{\em IEEE Trans. Inform. Theory}, vol. 34, no. 3, May 1988, pp. 380--388.
	



	\bibitem{Brandenburg:74}
	L. H. Brandenburg and A. D. Wyner.
	\newblock{``Capacity of the Gaussian channel with memory: The multivariate case,"}
	\newblock{\em Bell System Technical Journal}, vol. 53, no. 5, May. 1974,  pp. 745-778.
	
	

	\bibitem{Verdu:88}
	S. Verdu.
	\newblock{``The capacity region of the symbol-asynchronous Gaussian multiple-access channel,"}
	\newblock{\em IEEE Trans. Inform. Theory}, vol. 35, no. 4, Aug. 1989, pp. 733--751.	


	\bibitem{Goldsmith:05}
	A. Goldsmith.
	\newblock{\em Wireless Communications}.
	\newblock Cambridge, 2005.








	

%
	
	

	\bibitem{Lin:13}
	J. Lin, M. Nassar, and B. L. Evans.
	\newblock{``Impulsive noise mitigation in powerline	communications using sparse Bayesian learning,"}
	\newblock{\em IEEE J. Sel. A. Commun.}, vol. 31, no. 7, Jul. 2013, pp. 1172--1183.
	
	\bibitem{Lampe:16}
	L. Lampe, A. M. Tonello, and T. G. Swart.
	\newblock{\em Power line communications: Principles, standards and applications from multimedia to smart grid}.
	\newblock Wiley press, 2016.

	\bibitem{Giannakis:98}
	G. B. Giannakis.
	\newblock{``Cyclostationary signal analysis,"}
	\newblock{\em Digital Signal Processing Handbook}, CRC Press, 1998, pp. 17.1--17.31.
	
	\bibitem{Schrempf:05}
	O. C. Schrempf,  O. Feiermann, and  U.D. Hanebeck.
	\newblock{``Optimal mixture approximation of the product of mixtures,"}
	\newblock{\em IEEE International Conference on Information Fusion}, Philadelphia, PA, Jul. 2005.


	\bibitem{Tse:05}
	D. Tse and P. Viswanath.
	\newblock{\em Fundamentals of  Wireless Communication}.
	\newblock Cambridge, 2005.
	
	\bibitem{ITU:11}
	International Telecommunications Union (ITU).
	\newblock{``ITU-T Recommendation G.9963, Unified high-speed wire-line based home networking transceivers — Multiple Input/Multiple Output (MIMO),"}
	\newblock{\em } Sep. 2011.
	


	\bibitem{Evans:12}
	M. Nassar, J. Lin, Y. Mortazavi, A. Dabak, I. H. Kim, and B. L. Evans.
	\newblock{``Local utility power line communications in the 3--500 kHz band: Channel impairments, noise, and standards,"}
	\newblock{\em IEEE Signal Processing Magazine}, vol. 29, no. 5, Aug. 2012, pp. 116-127.
	
	\bibitem{IEEE:11}
	\newblock{\em Appendix for Noise Channel Modeling for IEEE P1901.2},
	\newblock{\em } IEEE Standard P1901.2, Jun. 2011.
	

	\bibitem{Amann:05}		
	H. Amann and J. Escher.
	\newblock{\em Analysis I}.
	\newblock Birkhauser Verlag, Basel, 2005.
	
	\bibitem{Guttierez:08}
	J. Gutti{\'e}rez-Gutti{\'e}rez and P. M. Crespo.
	\newblock{``Asymptotically equivalent sequences of matrices and	hermitian block toeplitz matrices with continuous symbols: Applications to MIMO systems,"}
	\newblock{\em IEEE Trans. Inform. Theory}, vol. 54, no. 12, Dec. 2008, pp. 5671--5680.

	\bibitem{Huber:08}
	M. F. Huber, T. Bailey, H. Durrant-Whyte, and U. D. Hanebeck.
	\newblock{``On entropy approximation for Gaussian mixture random vectors,"}
	\newblock{\em IEEE International Conference on Multisensor Fusion and Integration for Intelligent Systems (MFI)}, Seoul, South Korea, Aug. 2008, pp. 181-188.
	
	\bibitem{Meyer:00}
	C. D. Meyer.
	\newblock{\em Matrix Analysis and Applied Linear Algebra}.
	\newblock Society for Industrial and Applied Mathematics, 2000.
	
	\bibitem{Papoulis:91}
	A. Papoulis.
	\newblock{\em Probability, Random Variables, and Stochastic Processes}.
	\newblock McGraw-Hill, 1991.
	

	\bibitem{Michalowicz:13}
	J. V. Michalowicz and J. M. Nichols.
	\newblock{\em Handbook of differential entropy}.
	\newblock CRC press, 2013.

	
	
	\bibitem{Shannon:48}
	C. E. Shannon.
	\newblock{``A mathematical theory of communication,"}
	\newblock{\em Bell System Technical Journal}, vol. 27, no. 3-4, Jul./Oct. 1948,  pp. 379–-423, 623–-656.



    \bibitem{DiggaviCover:2001}
    S. N. Diggavi and T. M. Cover.
    \newblock{``The worst additive noise under a covariance constraint,"}
    \newblock{\em IEEE Trans. Inform. Theory}, vol. 47, no. 7, Dec. 2001, pp. 3072--3081.

    	
	\bibitem{Shlezinger:16b}
	N. Shlezinger and R. Dabora.
	\newblock{``The capacity of discrete-time Gaussian MIMO channels with periodic characteristics,"}
	\newblock{\em IEEE International Symposium on Information Theory (ISIT)}, Barcelona, Spain, Jun. 2016.




	\bibitem{IEEE:10}
	\newblock{\em IEEE Standard for Broadband over Power Line	Networks: Medium Access Control and Physical
		Layer Specifications},
	\newblock{\em } IEEE Standard P1901-2010, Dec. 2010.




}
\end{thebibliography}
\end{document}